\documentclass[%
 aip,
rsi,%
amsmath,amssymb,
preprint,%
]{revtex4-1}

\usepackage{graphicx}
\usepackage{dcolumn}
\usepackage{bm}
\usepackage{color}

\usepackage{mathtools}

\usepackage[version=3]{mhchem} 
\usepackage[T1]{fontenc}       
\usepackage{amsmath}
\usepackage{braket}
\usepackage{multirow}
\usepackage{bm}
\usepackage{dcolumn}
\usepackage{subcaption}
\usepackage{booktabs}
\usepackage[round-mode=places,round-precision=2]{siunitx}
\usepackage{blkarray}
\usepackage{morefloats}
\usepackage{float}
\usepackage[utf8]{inputenc}
\usepackage{soul}

\newcolumntype{d}{D{.}{.}{6}}

\begin{document}

\title[]{Vibrationally Resolved Coupled Cluster X-Ray Absorption Spectra from Vibrational Configuration Interaction Anharmonic Calculations} 

\author{Torsha Moitra}
\affiliation{DTU Chemistry - Department of Chemistry, Technical University of Denmark, Kemitorvet Bldg 207, DK-2800 Kongens Lyngby, Denmark}
\author{Diana Madsen}
\affiliation{Department of Chemistry, Aarhus University,   Langelandsgade~140, DK-8000 Aarhus C, Denmark}
\author{Ove Christiansen}
\email{ove@chem.au.dk}
\affiliation{Department of Chemistry, Aarhus University, Langelandsgade~140, DK-8000 Aarhus C, Denmark}
\author{Sonia Coriani}%
\email{soco@kemi.dtu.dk}
\affiliation{DTU Chemistry - Department of Chemistry, Technical University of Denmark, Kemitorvet Bldg 207, DK-2800 Kongens Lyngby, Denmark}

\date{\today}

\begin{abstract}
Vibrationally resolved near-edge x-ray absorption spectra at the K-edge for a number of small molecules have been computed from anharmonic vibrational 
configuration interaction
calculations of the Franck-Condon factors. The potential energy surfaces for ground and core-excited states were obtained at the core-valence separated CC2, CCSD, CCSDR(3) 
and CC3 levels of theory, employing the 
Adaptive Density-Guided Approach (ADGA) scheme
to select the single points at which to perform the energy calculations.
{We put forward an initial attempt to include pair-mode coupling terms to describe the potential of polyatomic molecules.}
%

\end{abstract}

\pacs{Valid PACS appear here}
\maketitle

\section{Introduction}
High-resolution inner-shell spectroscopies of molecules, in particular using soft-x-ray beamlines at the third-generation synchrotron radiation sources,~\cite{Ueda_2003} yield x-ray absorption (XAS) spectral bands with clearly identifiable vibrational structures.
Numerous experimental studies have been presented on a variety of small to medium size molecules over the last two decades~\cite{remmers1992high,Kempgens_1997,Prince_1999,DeSimone_2001,Prince_2002,Prince_small_2003,Hergenhahn_2004,Adachi_2005} in both photoabsorption and photoionization regimes.
Their results represent ideal cases to test the performance of computational methodologies to accurately reproduce and assign the individual spectral features.
From the vibrational structure of the spectrum,
one can for instance evaluate the shape of the potential energy surface (PES) and determine the effect of vibrations on the excitation energy.

Vibrationally resolved XAS spectra have been obtained before at the symmetry-adapted cluster–configuration interaction (SAC–CI),~\cite{Ehara_2008,Ehara_2010,Ehara_2011} algebraic diagrammatic construction (ADC)~\cite{cvs-adc-3} and time-dependent density functional (TD-DFT) levels of theory. As the applicability of coupled cluster (CC) response methods to x-ray spectroscopy has been significantly extended in recent years,~\cite{coriani2012pra,coriani2012jctc,MRCC_Core,Peng2015,coriani_cvs:2015,coriani_cvs:2016,naturecomm,Myhre_N2_2018,2019:Vidal:JCTC} it is interesting to investigate their accuracy in yielding vibrationally resolved XAS spectra. 
A detailed benchmark has recently been published~\cite{Myhre_N2_2018} on the N$_2$ molecule, where various CC approximations up to inclusion of full quadruple excitations were analyzed, and the core-valence separation~\cite{coriani_cvs:2015} (CVS) was exploited to obtain the core excited states.
A simulation of the vibronic progressions in the 1s$\to \pi^*$ bands of both carbon and oxygen K-edges in formaldehyde has also recently been performed at the CVS-CCSD
level~\cite{federica2019}
in the harmonic approximation.  
The  PES  of  the ground  state and of the  core excited  states  were  modeled  with  a  quadratic  expansion  around  their  minima  (i.e. with  the adiabatic Hessian  approach) and the  absorption  spectra  computed  in  Franck-Condon (FC) approximation, adopting  both  a time-independent and  the  time-dependent method.
Various other methods are documented in literature to compute the absorption spectra beyond the FC approximation. \cite{Non-condon_OPA_GFP_Anna_jpcl2011,BeyondFC_jpcl2018,vibrationallyCoupled_DuschinskyEffect_JCTC2020}

We use here the CC methods CC2 (coupled-cluster singles and approximate doubles\cite{cc2_method}), 
CCSD (coupled-cluster singles and doubles~\cite{CCSD,ove.1996.singlet}), 
CCSDR(3) and 
CC3 (coupled-cluster singles, doubles and
approximate triples~\cite{christiansen_perturbative_1996,CC3,CC3resp}), and vibrational configuration interaction (VCI[$n$]) to obtain the 
Franck-Condon factors.
The  Adaptive Density-Guided Approach (ADGA) scheme is applied to
automatically construct the 
necessary PES surfaces. 

ADGA employs densities from vibrational structure calculations for a dynamic sampling of PESs~\cite{Sparta2009_ADGA}
and has recently been extended to a multi-state version.\cite{msadga}
{We compare simulated and experimentally 
observed spectra and discuss the possibilities of sufficient spectral simulation using low-level coupling or even uncoupled PES representations.  
}

\section{Computational methodology}

\subsection{CVS-CC excitation energies}
Coupled cluster methods are built upon an exponential ansatz of the wavefunction, 
\begin{align}
    |\textrm{CC}\rangle = \exp{{T}}|\textrm{HF}\rangle
\end{align}
where $|\textrm{HF}\rangle$ is the reference Hartree-Fock wavefunction and $T=\sum_\mu t_\mu \tau_\mu$ is the cluster operator, with $t_\mu$ being the cluster amplitudes and $\tau_\mu$ being the corresponding excitation operators.

The ground state energy and amplitudes are typically determined by projection of the Schr{\"o}dinger equation onto the reference state, and onto a manifold of excitations $|\mu\rangle$ out of the reference state, respectively~\cite{helgaker2004}
\begin{align}
E &= \langle{\rm{HF}} | \exp (-T) H \exp (T)|{\rm{HF}}\rangle; \quad\quad
\\\nonumber
\Omega_\mu   &= \langle \mu | \exp(-T) H \exp(T) | {\rm{HF}}\rangle = 0
\end{align}
Within CCLR theory,~\cite{CC:RSP:1990,ove.1996.singlet,ijqc-ove} excitation energies ($\omega_k$) and left (${{L}}_k$) and right (${{R}}_k$) excitation vectors can be obtained by solving the asymmetric eigenvalue equations
\begin{align}
\label{eigenvalue_eq}
{\bf{A}} {{R}}_k = \omega_k {{R}}_k; \quad {{L}}_k{\bf{A}}  = \omega_k {{L}}_k
\end{align}
under the biorthogonality condition ${{L}}_j {{R}}_k = \delta_{ik}$. The Jacobian matrix ${\bf{A}}$ is defined as
\begin{align}
A_{\mu\nu} = \frac{\partial \Omega_{\mu}}{\partial t_{\nu}} = 
\langle \mu | \exp(-T) [H,\tau_{\nu}] \exp(T) | {\rm{HF}} \rangle
\end{align}

The CC2,\cite{cc2_method} CCSD,~\cite{CCSD,ove.1996.singlet} CCSDR(3),~\cite{christiansen_perturbative_1996} and CC3~\cite{christiansen_perturbative_1996,CC3,CC3resp} methods adopted here for the ground and excited state potential energy surfaces are well known approximations based on the coupled cluster wave function ansatz. Therefore, we refrain from a detailed description of their parametrizations.~\cite{ijqc-ove}
Briefly, CCSD is defined truncating the operator $T$ to single  and double excitations. In CC2 and CC3, additional approximations are introduced in the $\Omega_\mu$ function for the doubles and triples, respectively, and, consequently in the Jacobian matrix, based on {arguments related to perturbation theory and response theory}. 
The CCSDR(3) approach was specifically designed to  non-iteratively correct for triple excitation effects in the calculation of excitation energies within CCSD.~\cite{christiansen_perturbative_1996}
{The total excited state energy is in this case obtained by adding the CCSDR(3) excitation energy to the CC(3)\cite{CC3} ground state energy. The later approach corresponds to one iteration in ground-state CC3 based on a CCSD starting point. Informally, CCSDR(3) is also a kind of first CC3 iteration to the excitation energy using a CCSD starting point. 
We note in passing that the CC(3) energy is extremely close to CCSD(T).}
As we are specifically targeting the potential energy surfaces of core-excited states, 
a core-valence separation  projector~\cite{coriani_cvs:2015,coriani_cvs:2016,cederbaum:cvs:1980}
has been applied during the calculation of the core 
excited state energies.
{The projector ensures that the core excited states are decoupled from the valence excited states
by removing pure valence excitations from the trial vectors during the iterative solution of the eigenvalue equations, Eq.~\ref{eigenvalue_eq}.}
{For CCSDR(3) and CC3, care has to be taken 
to only project out the contributions of the triples orbital energy differences $\Delta\epsilon_{\mu_3}$ in the Jacobian blocks originating from the target states and not from the ground state.}

\subsection{PES construction}

Given that the total molecular PES, $V(\mathbf{q})$, depends on the full
set of $M = 3N_{\mathrm{nuc}}-6(5)$ vibrational coordinates or modes,
$\mathbf{q} = \{q_1, q_2, \dots, q_M\}$, its exact construction is impossible,
even for very approximate electronic structure. 
Instead, we here prioritize a high electronic structure treatment to describe
the basic process as outlined in the previous section and then we adjust the PES description to give a level of treatment that is sufficient to capture the trends in the vibrational structure. 

To do so, we work within an approximate scheme of systematic mode-coupling 
expansion, also denoted as n-mode PESs, 
where the PES 
is written as 
\begin{equation} \label{eq:n_mode_exp_short}
    V(\mathbf{q}) = \bar{V}^{0} + \sum_{i=1}^{M} \bar{V}^{m_i} + \sum_{j>i=1}^{M} \bar{V}^{m_i m_j} + \cdots.
\end{equation}
Here, the constant $\bar{V}^{0}$ is a reference point, which is usually chosen to
be the absolute minimum of the PES,
even though this is not a requirement.
The bar-potentials 
    $\bar{V}^{m_i} \equiv V^{m_i} - \bar{V}^{0},  \label{eq:pot_bar_func_2}$
    $\bar{V}^{m_i m_j} \equiv V^{m_i m_j} - \bar{V}^{m_i} - \bar{V}^{m_j} + \bar{V}^{0}$, 
    etc., 
are generated from the cut-potentials of the form
    $V^{m_i} \equiv V(0,\dots,0, q_{m_i} ,0,\dots,0)$ and 
    $V^{m_i m_j} \equiv V(0,\dots,0,q_{m_i},0,\dots,0, q_{m_j},0,\dots,0)$.
Inclusion of only up to the one-mode terms in Eq.~\eqref{eq:n_mode_exp_short} defines
what we throughout will denote a 1M PES. 
Including all terms up to the two-mode level defines what we will denote as 2M PES. 
The exact PES is obtained only when the expansion order n is equal to $M$. 
We may even restrict these expansions by including only specific modes or specific couplings. 
{We also note that, 
while truncating
already at the 1M level or 
restricting the 
couplings to only a few modes 
may not be sufficient for high precision studies of frequencies as measured in IR, 
one should acknowledge that  that primary target here
is the vibrational profile of an x-ray absorption spectrum, 
and the ability to simulate such spectral profiles across several states and molecules.
}


The specific challenge in  computations of FC profiles is that two electronic states are involved, 
and what is the optimal set of vibrational coordinates of these may differ. 
We take here the approach of using one fixed set of coordinates for all computations,
and then expand the PES for that set using the different states and different
electronic structure methods. 
Then, the level of mode-coupling will be pragmatically investigated until the main features
have been identified.
{Correspondingly, we use multi-state ADGA with tight threshold to ensure convergence. 
ADGA computations are driven by the average vibrational state density as an importance weighting factor, see Ref.\citenum{msadga} for more details. See also Ref.\citenum{madsen_vibrationally_2019} for a related application to the computation of FC profiles. }

\subsection{Anharmonic Franck-Condon factors from VCI wavefunctions}

{Within the selected vibrational subspace, we perform FC factor calculations based on VCI wavefunctions.\cite{msadga,madsen_vibrationally_2019} 
We refer to Refs.\citenum{rauhut_anharmonic_2015, rodriguez-garcia_franck-condon_2006, luis_variational_2006,bowman_determination_2006} 
for application of VCI FC factors in other contexts. VCI} introduces mode coupling as a linear expansion of excitations from a reference state. A VSCF wavefunction was used as the reference state for the VCI wavefunctions. A key feature of VCI wavefunction is the inclusion of mode coupling {in the wave function in a hierarchical manner},\cite{Ove_PCCP_2012} denoted as VCI[$n$], where $n$ is the maximum number of simultaneously excited modes.
The absorption spectra are calculated from the FC factors based on the overlaps of the VCI wavefunctions of the lowest vibrational state in the ground (initial) state and the different vibrational states of the excited (final) electronic state.
For the purpose of analysis we also present, in several cases, VCI[1] spectra. 
These spectra are typically obtained with only 1M PES where there are no couplings between modes and VCI[1] gives exact states (within the given one-mode basis). Note, however, that the state space only includes one-mode excited states. This is therefore an extremely simple approach, here presented to see how far we get with a minimal description as a sum of independent anharmonic oscillators. 
{In principle, one could easily stay at this level and include combination bands by appropriate factorizations of these FC factors, but we simply obtained these combination bands in our higher level VCI[$n$] computations. }
{Here the space and therefore the spectra also include states that are excited in several modes simultaneously. {When couplings in the PES are included, 
{then} the energy, states, and FC factors for the one-mode excited states will also be different from the former VCI[1] results.}
}

\section{Computational details}

The reference starting structures were generated by optimizing the ground state geometry at MP2/cc-pCVTZ level of theory using the CFOUR program package.~\cite{cfour} Vibrational coordinates obtained from the ground state Hessian were used to span the PES for both the ground and the excited states. The iterative ADGA scheme,~\cite{Sparta2009_ADGA} as implemented in MidasCpp,~\cite{MidasCpp201940} was employed to choose the single points at which to calculate the core excitation energies. At each distorted geometry generated by the ADGA procedure, a single point energy calculation for the given CC method 
was performed by interfacing MidasCPP with the Dalton program.\cite{DaltonWIRES} The PES were fitted using polynomial fitting of maximum order 12. 

Mode coupling has been considered at two stages in our calculations: firstly, while generating of PES and, secondly, while performing the VCI calculations. 
Except for the case of formaldehyde and acetaldehyde, 1M PES calculations are presented here.
To ensure that most of the intensity was captured in the simulated spectra, the sum of Franck-Condon factors was required to be greater than 0.98. 
{For {all} molecules 
VCI[1] was used to compute the FC factors and is 
exact {within the given one mode basis}. 
{For the polyatomic
molecules},
we used {additionally} 
VCI[4] to calculate the FC factors.}

We will throughout denote the VCI 
calculations
as VCI[1] and VCI[4] for simplicity, even though in some cases the VCI[4] {analysis} was limited to two and three modes, and therefore they 
could be more {precisely}
denoted as VCI[2] and VCI[3] computations{, respectively.} 

In the case of molecules with multiple normal modes, we first carried out a calculation with all modes at the CCSD level using a smaller basis set (cc-pVDZ). Then, we only considered the major contributing modes in a more detailed study using the cc-pCVTZ basis set. For the case of OCS, we have relied on previous studies for the choice of contributing vibrational mode.\cite{doi:OCS-renner}

X-ray absorption spectra were generated based on the computed FC factors, using Lorentzian broadening, with HWHM of 0.04~eV. 
In all figures shown below, 
the calculated spectra (blue solid line) were scaled and shifted in order to match the highest experimental peak with the calculated one, unless otherwise mentioned. A red vertical line marks the position of the vertical electronic excitation shifted by the same amount.
Experimental spectra have been digitized from the original references using WebPlotDigitizer~\cite{WPD} and are plotted with green dotted lines. The orange sticks are the FC factors. 
In all the PES plots, the solid lines refer to the polynomial fitted PESs and dashed lines are the mean densities of states. 
\section{Results and Discussions}
\subsection{Carbon monoxide (CO)}
A highly resolved energy-loss spectrum at the carbon K-edge in CO, corresponding to the 1s excitation to the first unoccupied valence orbital $\pi^*$, was reported both by Tronc \textit{et al.}\cite{Tronc_1979} and by \citeauthor{Maexp}\cite{Maexp}
The experimental spectrum exhibits vibrational progression with the main peak at 287.40 eV attributed to the transition to the $v=0$ level of the $\pi^*$ state. The less intense peaks at 287.66 eV and 287.91 eV were assigned to be transitions to the $v=1$ and $v=2$ states, with intensities less than the first peak by 
an order of magnitude of one and two, respectively.
The oscillator strength of the first peak is much higher than the rest,
{in line with the more compact nature of the $\pi^*$ orbital}. 
\begin{figure}[htbp]
\centering
    \begin{subfigure}[b]{0.4\textwidth}
    \includegraphics[scale=0.75]{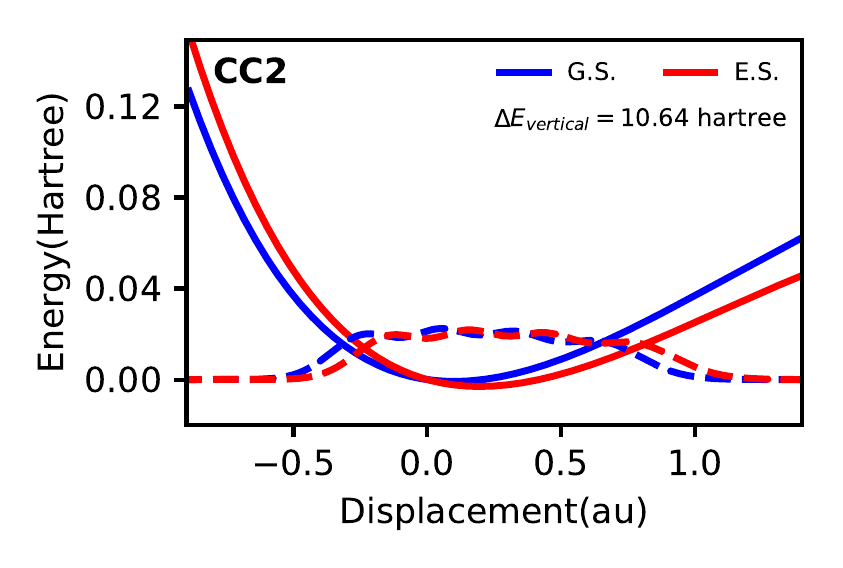}
    \end{subfigure}
     \begin{subfigure}[b]{0.4\textwidth}
    \includegraphics[scale=0.75]{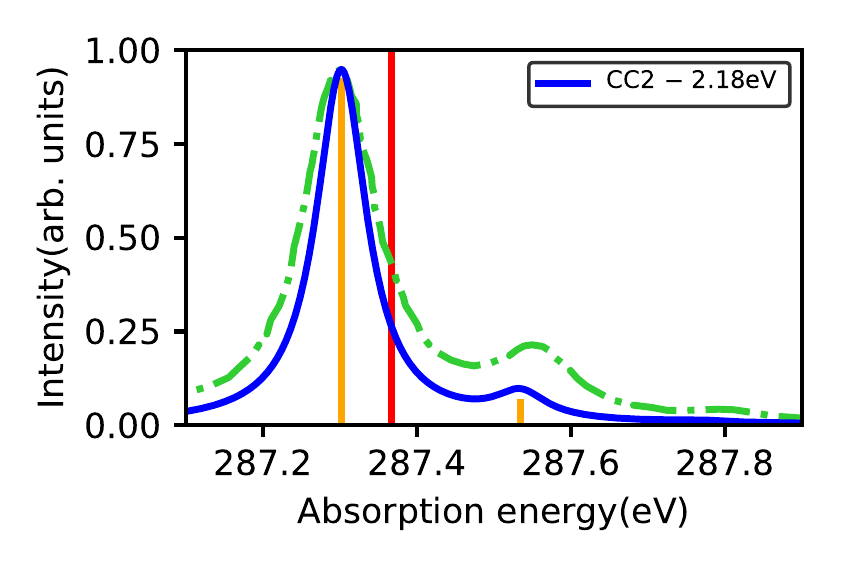}
    \end{subfigure}
    \begin{subfigure}[b]{0.4\textwidth}
    \includegraphics[scale=0.75]{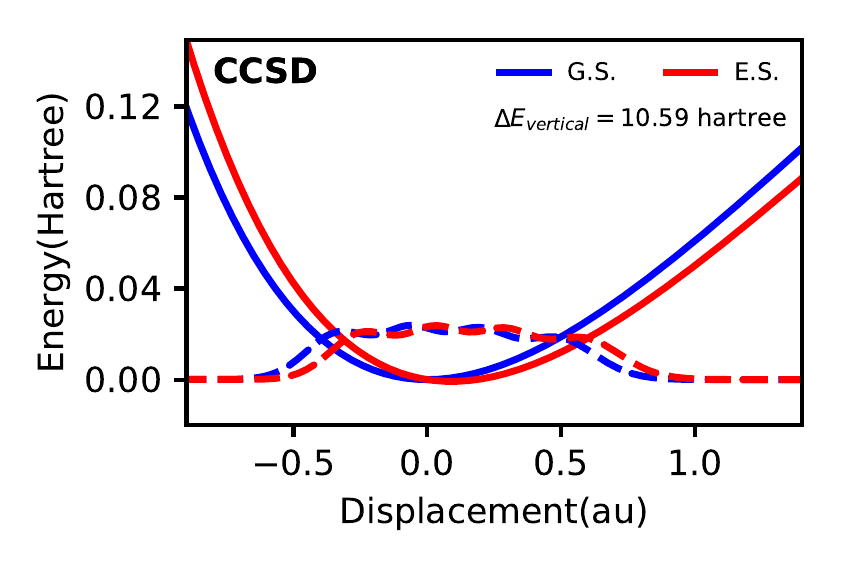}
    \end{subfigure}
    \begin{subfigure}[b]{0.4\textwidth}
    \includegraphics[scale=0.75]{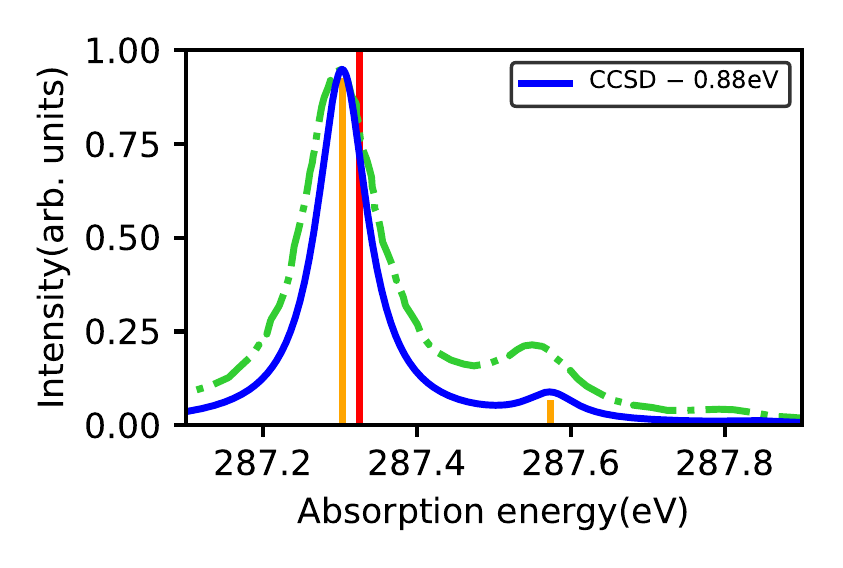}
    \end{subfigure}
    \begin{subfigure}[b]{0.4\textwidth}
    \includegraphics[scale=0.75]{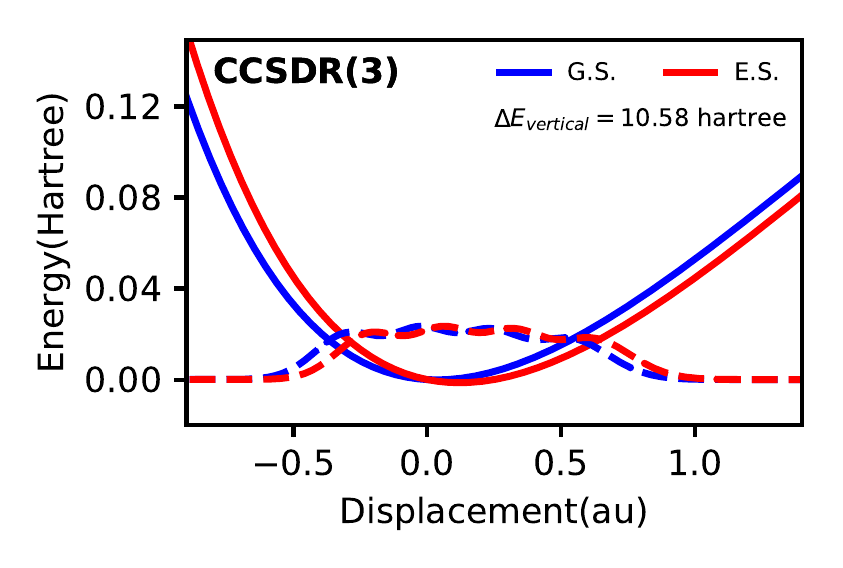}
    \end{subfigure}
    \begin{subfigure}[b]{0.4\textwidth}
    \includegraphics[scale=0.75]{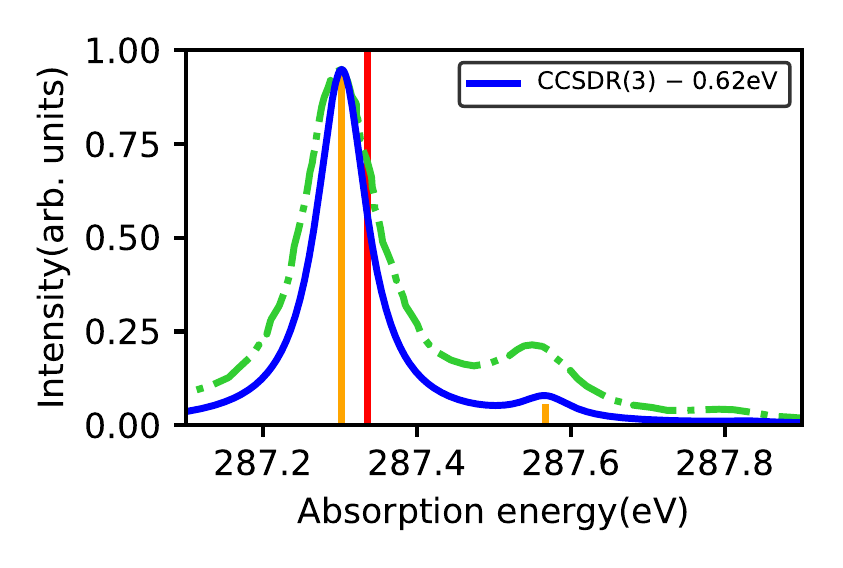}
    \end{subfigure}
    \begin{subfigure}[b]{0.4\textwidth}
    \includegraphics[scale=0.75]{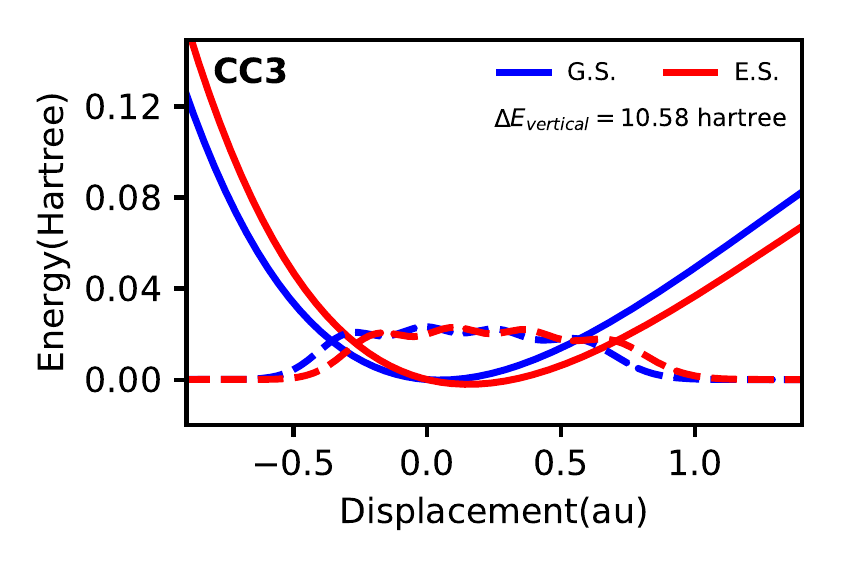}
    \end{subfigure}
    \begin{subfigure}[b]{0.4\textwidth}
    \includegraphics[scale=0.75]{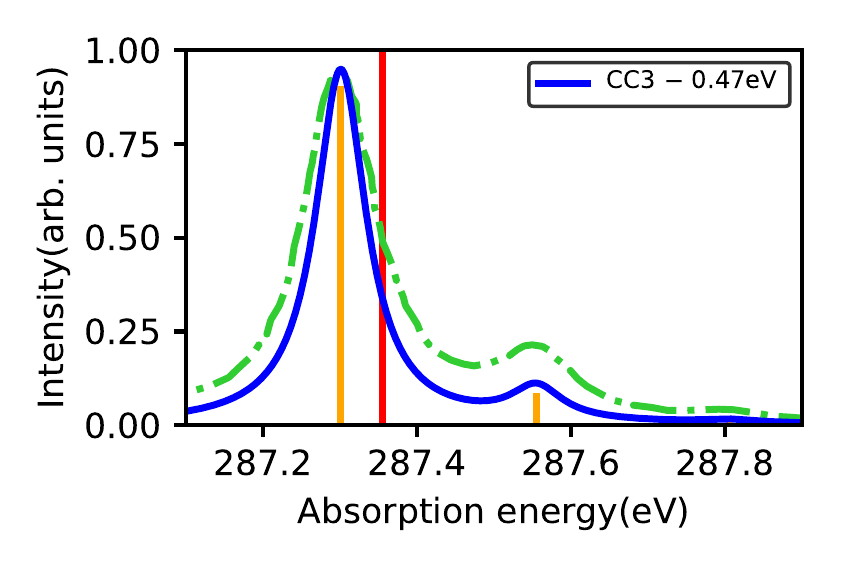}
    \end{subfigure}
    \caption{CO: 
    PESs (left) and vibrationally resolved 1s$_{\rm{C}}\to\pi^*$ spectral band (right).
    PESs are shown as solid lines and mean densities of the states as broken lines. Blue and red color correspond the ground and the excited state, respectively. For the purpose of comparison a common zero at the reference point has been used, i.e. the excited state is shifted down to the ground state. In the spectra (right), the orange vertical lines correspond to the FC factors. The experimental spectrum is shown as green dotted line. The shift applied to align to the experimental first peak is given in legend. A red vertical line marks the position of the vertical excitation energy, shifted by the same amount.}
    \label{fig:C_CO}
\end{figure}
Our results for the vibrationally resolved 1s$_{\rm{C}} \to \pi^*$ band are presented in the right panels of Fig.~\ref{fig:C_CO}.
They were obtained from the FC factors calculated using CC PESs and vibrational mean densities of the ground and core-excited electronic states as a function of the nuclear displacement shown in the left panels of Fig.~\ref{fig:C_CO}.

All our calculations reproduce the shape of the vibrational progression $v=0, v=1$ reasonably well, as shown in Fig.~\ref{fig:C_CO}. At all CC levels, the $v=2$ FC factor is obtained to be very low compared to $v=0$ and $v=1$ transitions and is thus not visible to the naked eye.

Different rigid shifts are  applied for the different CC methods to align with the experimental spectrum. The size of the shift applied decreases with increasing quality of the CC approach used.
The PES generated at CC2 level is the shallowest and, remarkably, the energy gap between the vibrational progression is best described at this level.
The CCSD PESs 
of the ground and excited state are the least shallow and result in an overestimation of the separation between the peaks. 
The CCSDR(3) and CC3 ones
become progressively shallower, correcting for the overestimated peak separation  at CCSD level. The ratios between the FC factors also improve going from CCSD to CC3.

The next electronic transition in the experiment is attributed to 
1s$_{\rm{C}}\to\sigma$3s Rydberg transition, with 3 vibrational peaks at 292.37, 292.65 and 292.93 eV.\cite{Tronc_1979} 
Simulating the spectral features of transitions to Rydberg states has some additional complications. Firstly, we had to augment the basis set with specialised Rydberg-type functions, whose exponents were computed according to the prescription of Kaufmann \textit{et al} (and with quantum number $n=3, 3.5, 4$).\cite{Kaufmann_1989}
Secondly, since this is not the lowest-energy transition, more roots were needed to obtain it. Thirdly, as the 1s$_{\rm{C}}$ $\to$3p Rydberg transition lies close to the 1s$_{\rm{C}}\to$3s Rydberg transition, a continuous monitoring had to be done to ensure that the correct state was being tracked.   

\begin{figure}[htbp]
    \centering
    \begin{subfigure}[b]{0.4\textwidth}
    \includegraphics[scale=0.8]{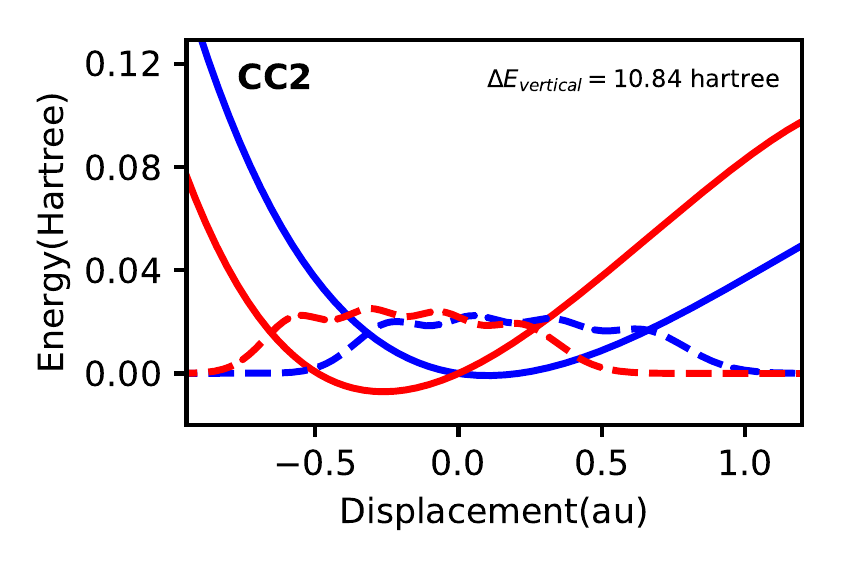}
    \end{subfigure}
    \begin{subfigure}[b]{0.4\textwidth}
    \includegraphics[scale=0.8]{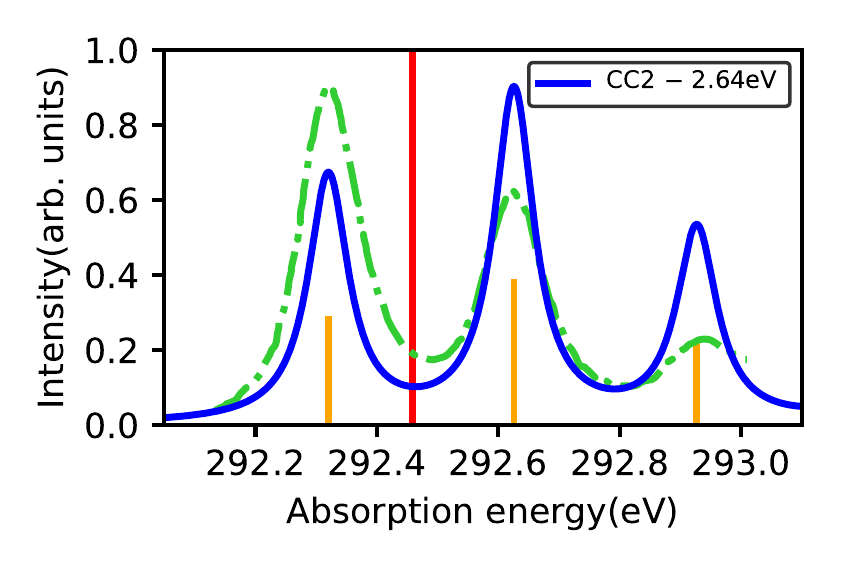}
    \end{subfigure}
    \begin{subfigure}[b]{0.4\textwidth}
    \includegraphics[scale=0.8]{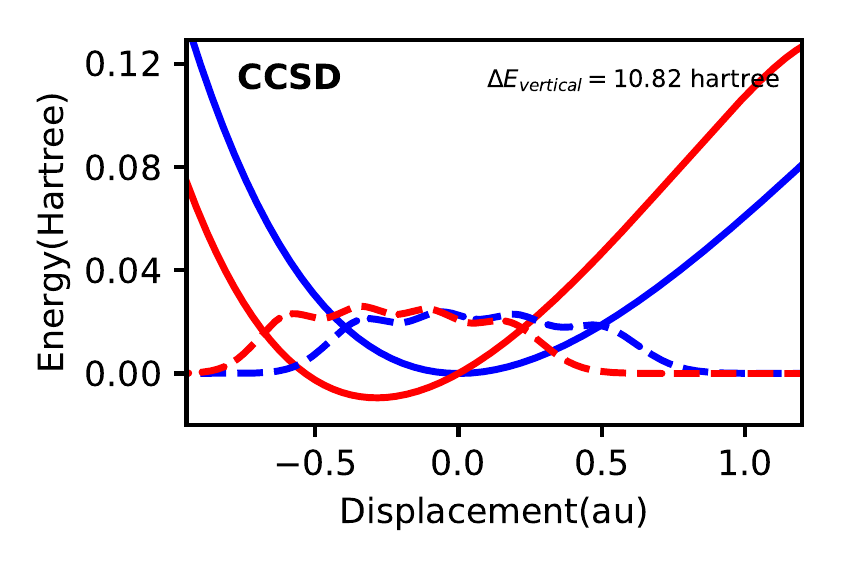}
    \end{subfigure}
    \begin{subfigure}[b]{0.4\textwidth}
    \includegraphics[scale=0.8]{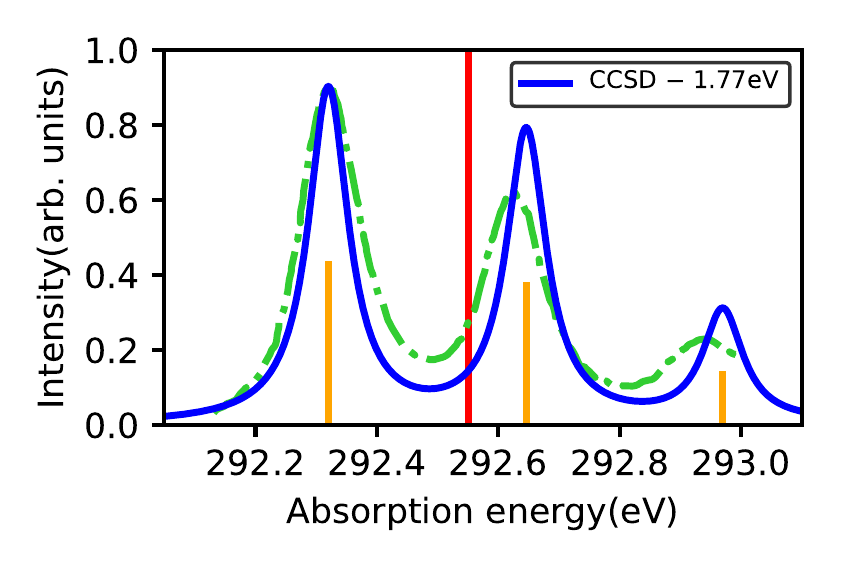}
    \end{subfigure}
    \begin{subfigure}[b]{0.4\textwidth}
    \includegraphics[scale=0.8]{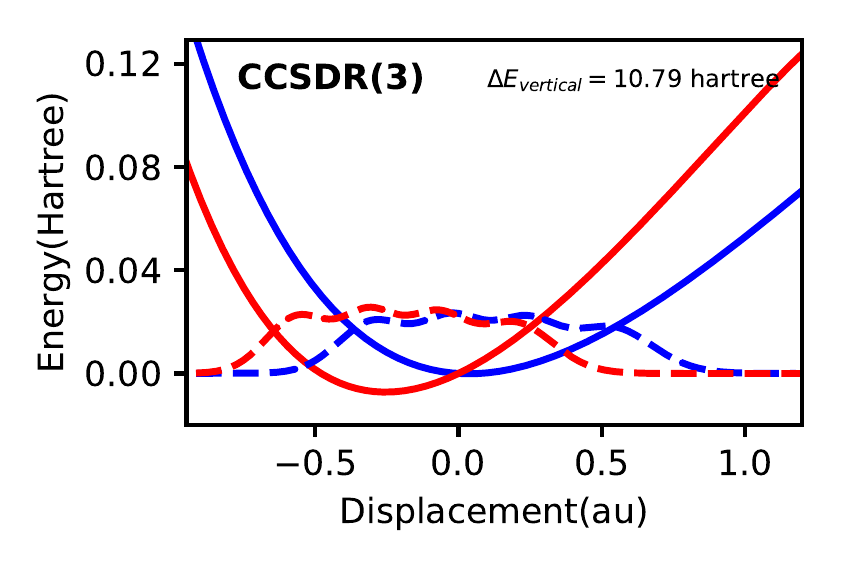}
    \end{subfigure}
    \begin{subfigure}[b]{0.4\textwidth}
    \includegraphics[scale=0.8]{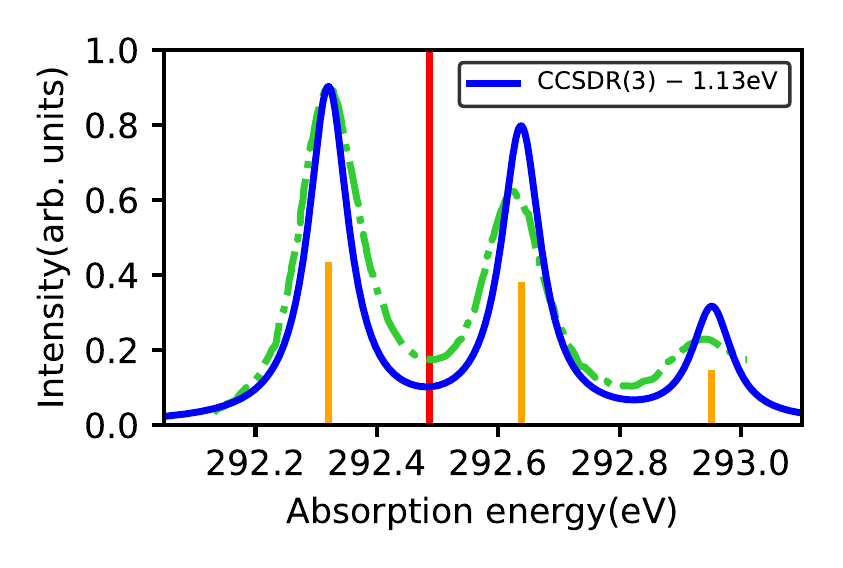}
    \end{subfigure}
     \caption{CO: PES and vibrationally resolved 1s$_{\rm C}$ $\rightarrow \sigma$3s Rydberg band. See the Computational Details section and the caption of Fig.~\ref{fig:C_CO} for an explanation of the notation.}
   \label{fig:CO-C-rydberg}
\end{figure}

The PES and the vibrationally resolved spectra we obtained for this electronic transition at the various levels of theory are shown in Fig.~\ref{fig:CO-C-rydberg}.
For all methods, three FC sticks were obtained.
At the CC2 level, the PES generated is shallower compared to those at the higher level theory. The energy separations between the peaks are, at CC2 level, rather accurate whereas the relative intensities between the vibrational progressions are clearly not consistent with experiment. {When plotting the spectra for the CC2 method, we align the position of the simulated 0-0 band to match the experimental first peak. }
The intensity of the first overtone is  overestimated in comparison to experiment by all three CC methods here considered.
The peak separation at the CCSD level is also slightly worse compared to experiment than the one yielded by the other methods.
We did not perform CC3 calculations as it became very expensive to compute multiple roots at multiple points.

\begin{figure}[htb!]
    \centering
    \begin{subfigure}[b]{0.4\textwidth}
    \centering
    \includegraphics[scale=0.8]{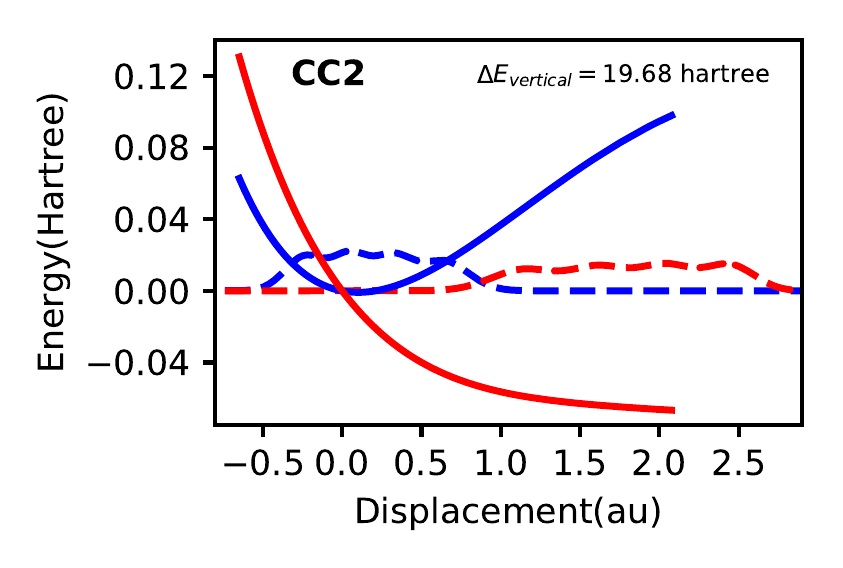}
    \end{subfigure}
    \begin{subfigure}[b]{0.4\textwidth}
    \centering
    \includegraphics[scale=0.8]{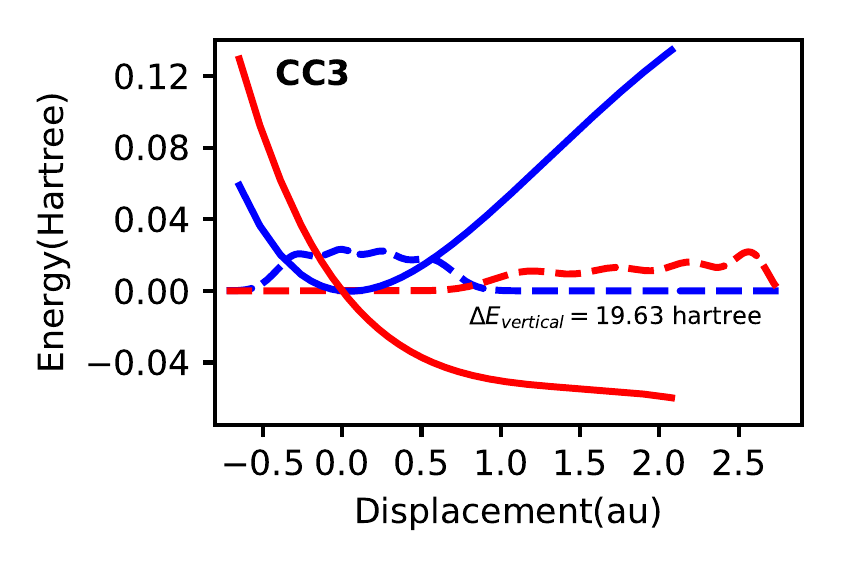}
    \end{subfigure}
    \begin{subfigure}[b]{0.4\textwidth}
    \centering
    \includegraphics[scale=0.8]{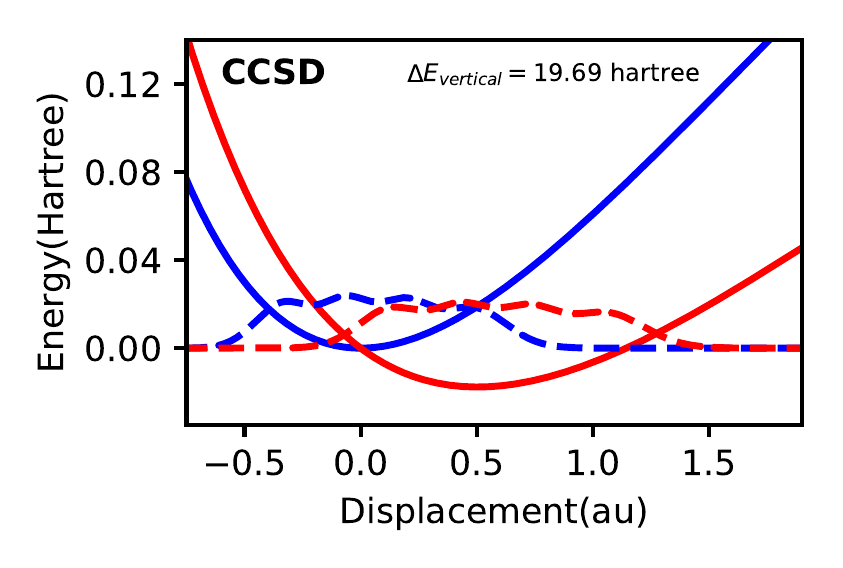}
    \end{subfigure}
    \begin{subfigure}[b]{0.4\textwidth}
    \centering
    \includegraphics[scale=0.8]{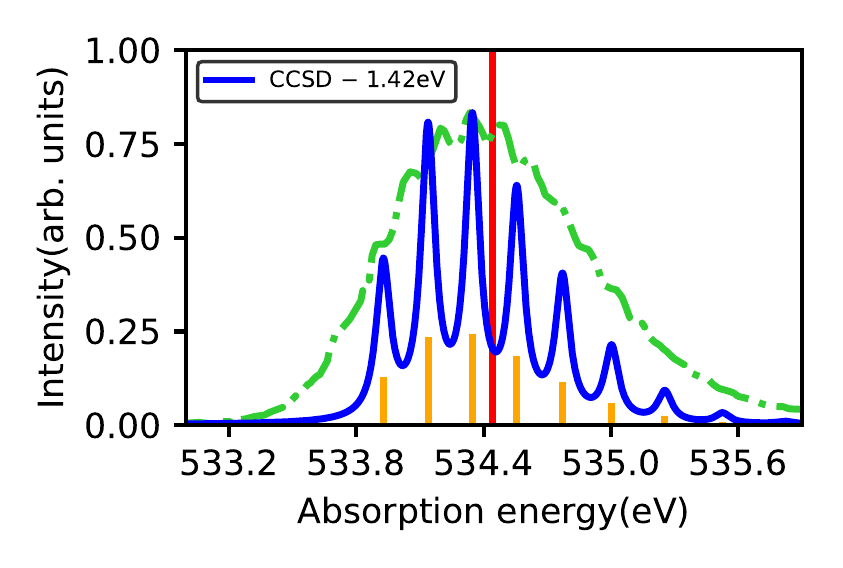}
    \end{subfigure}
    \begin{subfigure}[b]{0.4\textwidth}
    \centering
    \includegraphics[scale=0.8]{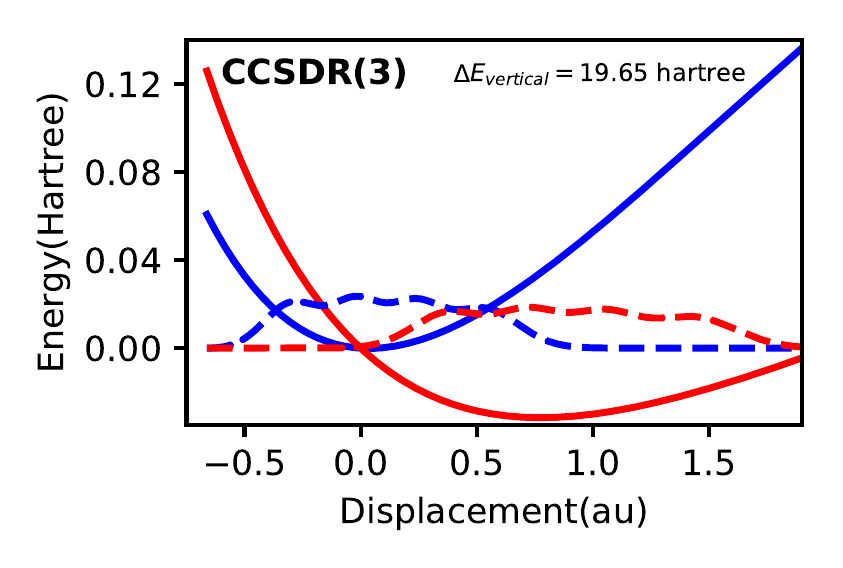}
    \end{subfigure}
    \begin{subfigure}[b]{0.4\textwidth}
    \centering
    \includegraphics[scale=0.8]{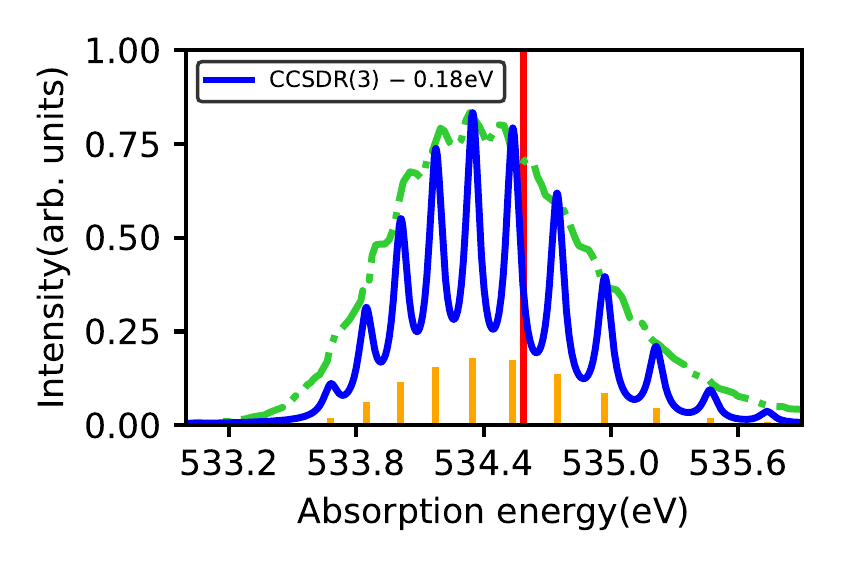}
    \end{subfigure}
\caption{CO: PES and vibrationally resolved band of the 1s$_{\rm{O}}$ $\to\pi^* $ transition. See the Computational Details section and the caption of Fig.~\ref{fig:C_CO} for an explanation of the notation.}
\label{fig:O_CO}
\end{figure}
We next turn our attention to the oxygen K-edge of CO. Vibrationally resolved XAS spectra of CO at the O K-edge were recorded by \citeauthor{PuttnerPRA}~\cite{PuttnerPRA} Though not very well resolved, the \mbox{1s$_{\rm O}\to\pi^*$} band shows a progression of eleven peaks. The most intense one was assigned to the 1s$(v=0)\to\pi^*(v=5)$ transition at 533.57 eV. 

As shown in the two upper panels of Fig.~\ref{fig:O_CO}, at both CC2 and CC3 levels, the PES of the 1s$_{\rm{O}}\to\pi^*$ core excited state were found to be dissociative. {Note that, in this case, ADGA, set out to compute bound states, did not converge. Therefore, the vibrational density profiles in  Fig.~\ref{fig:O_CO}
are rather artificial} and {only shown to
convey the information that
the potential {energy surfaces} {are} not converged to the dissociation limit 
{and thus unfit for FC analysis.}
}
The vibrational progression obtained at the CCSDR(3) level is better resolved than the one at CCSD level, with 11  peaks versus 9. The assignment of the most intense peak in the perturbative triples method matches the experimental observation. Additionally, 
the excited state PES is shallower and 
the mean excited-state vibrational density is more spread out over the displacement coordinate in CCSDR(3). 
The PES generated supports the observation made by \citeauthor{PuttnerPRA} \cite{PuttnerPRA} that 
the (equilibrium) bond distance increases upon core excitation to the $\pi^*$ orbital.

\begin{figure}[htbp]
    \centering
    \begin{subfigure}[b]{0.4\textwidth}
    \includegraphics[scale=0.8]{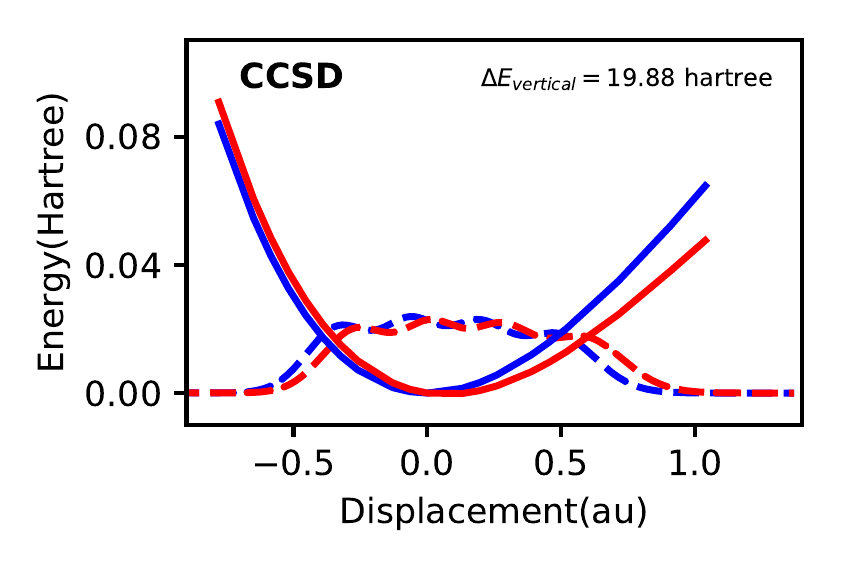}
    \end{subfigure}
    \begin{subfigure}[b]{0.4\textwidth}
    \includegraphics[scale=0.8]{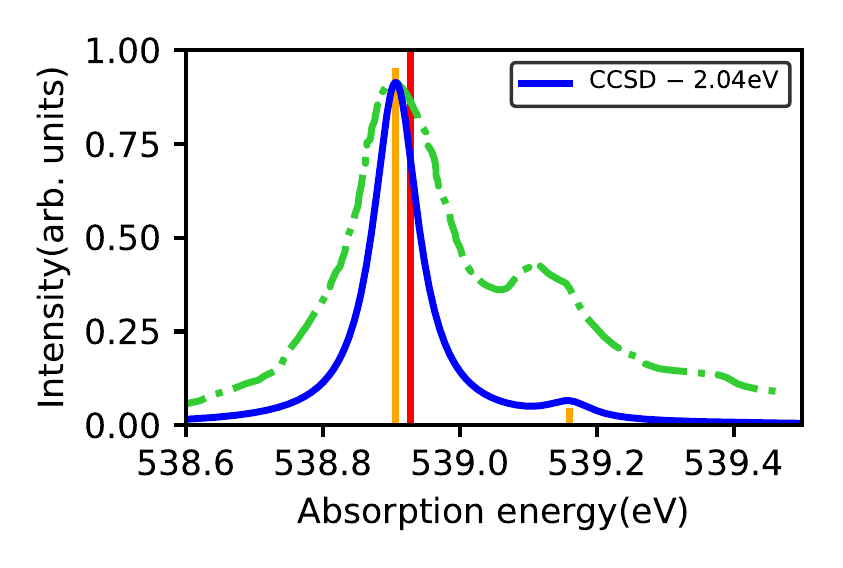}
    \end{subfigure}
    \begin{subfigure}[b]{0.4\textwidth}
    \includegraphics[scale=0.8]{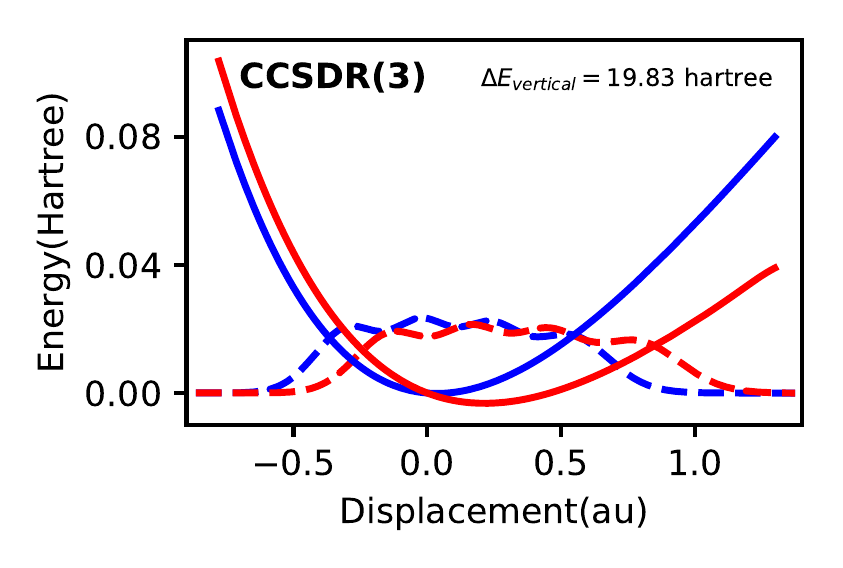}
    \end{subfigure}
    \begin{subfigure}[b]{0.4\textwidth}
    \includegraphics[scale=0.8]{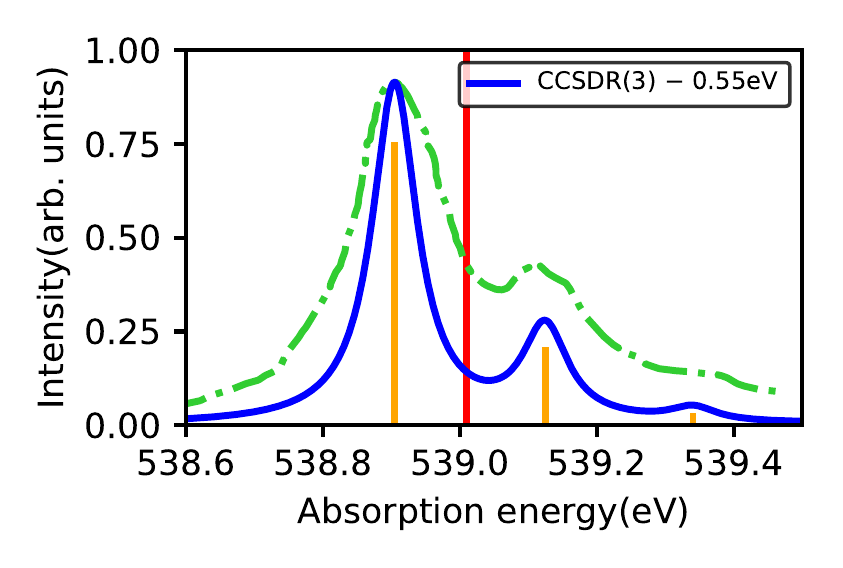}
    \end{subfigure}
\caption{CO: PES and vibrationally resolved XAS 1s$_{\rm O}\to$3s Rydberg band.
See the Computational Details section and the caption of Fig.~\ref{fig:C_CO} for an explanation of the notation.
}
\label{fig:O_CO_rydberg}
\end{figure}

We also explored the ADGA procedure to generate the PES for the 1s$_{\rm O}\to$3s Rydberg transition. The experimental spectrum shows 3 vibrational peaks at around 539~eV.\cite{PuttnerPRA} Also in this case we added Kaufmann's $n=3,3.5,4$ Rydberg functions to the basis set. Similar dissociative
character of the excited state PES from CC2 and CC3 as for the 1s$_{\rm{O}}$ $\to\pi^* $ band was observed, so these two methods have been omitted. On the other hand, stable CCSD and CCSDR(3) results were obtained and they are shown in Fig.~\ref{fig:O_CO_rydberg}. The CCSDR(3) PES  is once again shallower than in CCSD and the mean density of the state extends more towards a positive displacement. Remarkable differences in the computed XAS are observed, as CCSD fails to attain the lowest intensity third vibrational peak and significantly underestimates the intensity of the second peak. The perturbative triples correction helps in overcoming these shortcomings and yields a more accurate vibrationally resolved XAS band. 
\subsection{Nitrogen (N$_2$)}
\begin{figure}[htb!]
    \centering
    \begin{subfigure}[b]{0.4\textwidth}
    \centering
    \includegraphics[scale=0.8]{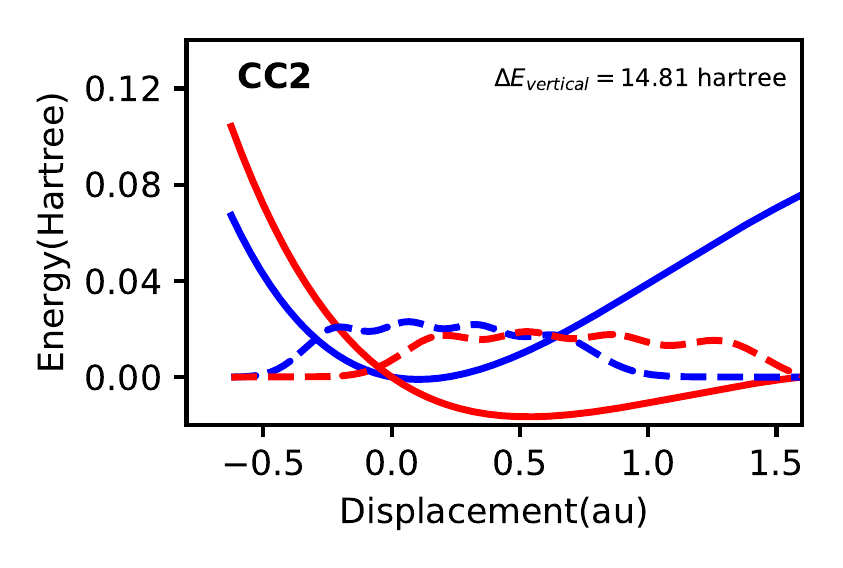}
    \end{subfigure}
    \begin{subfigure}[b]{0.4\textwidth}
    \includegraphics[scale=0.8]{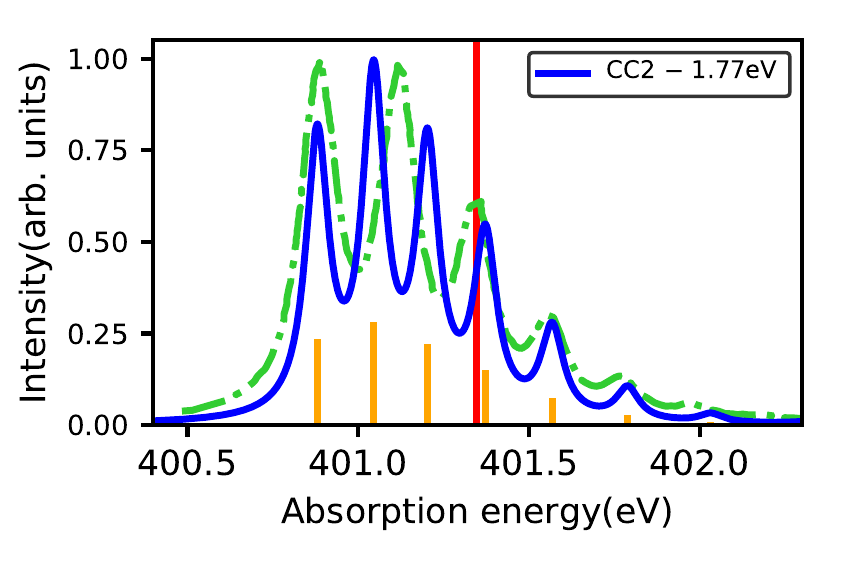}
    \end{subfigure}
    \begin{subfigure}[b]{0.4\textwidth}
    \centering
    \includegraphics[scale=0.8]{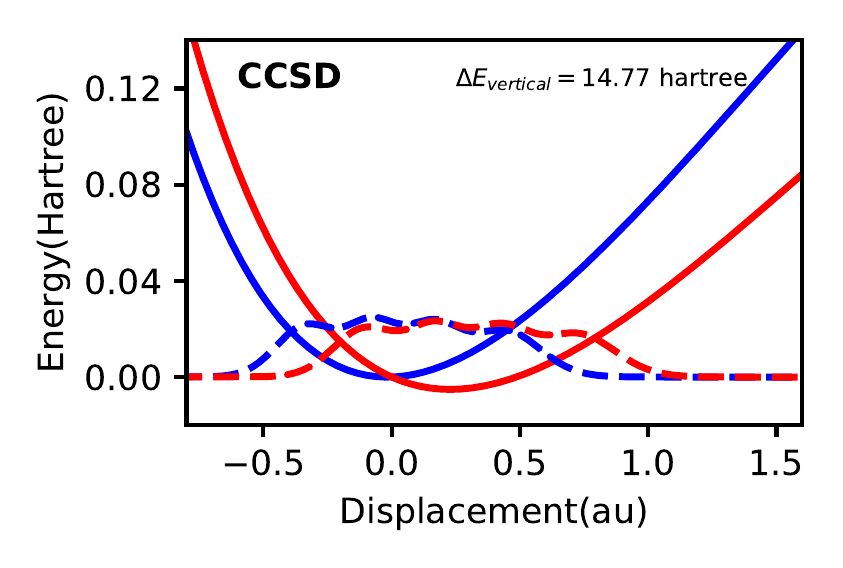}
    \end{subfigure}
    \begin{subfigure}[b]{0.4\textwidth}
    \includegraphics[scale=0.8]{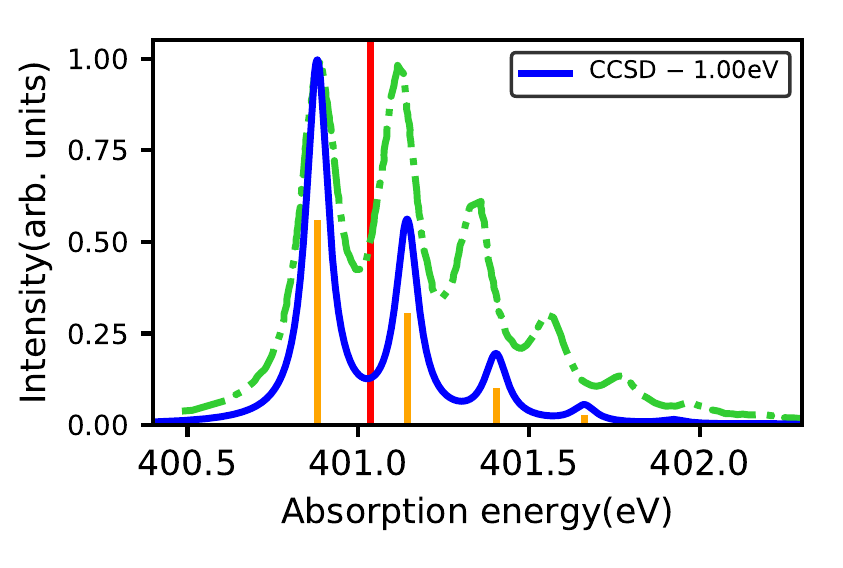}
    \end{subfigure}
     \begin{subfigure}[b]{0.4\textwidth}
        \centering
         \includegraphics[scale=0.8]{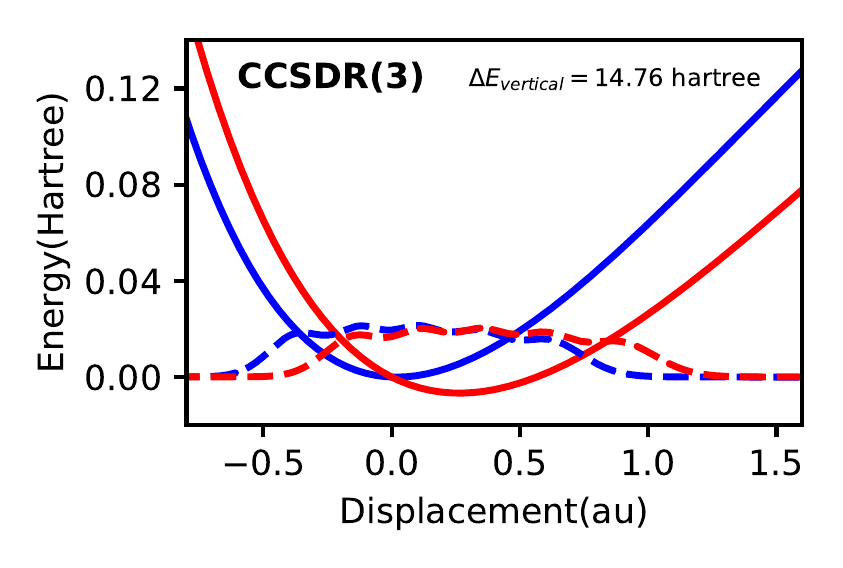}
     \end{subfigure}
     \begin{subfigure}[b]{0.4\textwidth}
      \centering
      \includegraphics[scale=0.8]{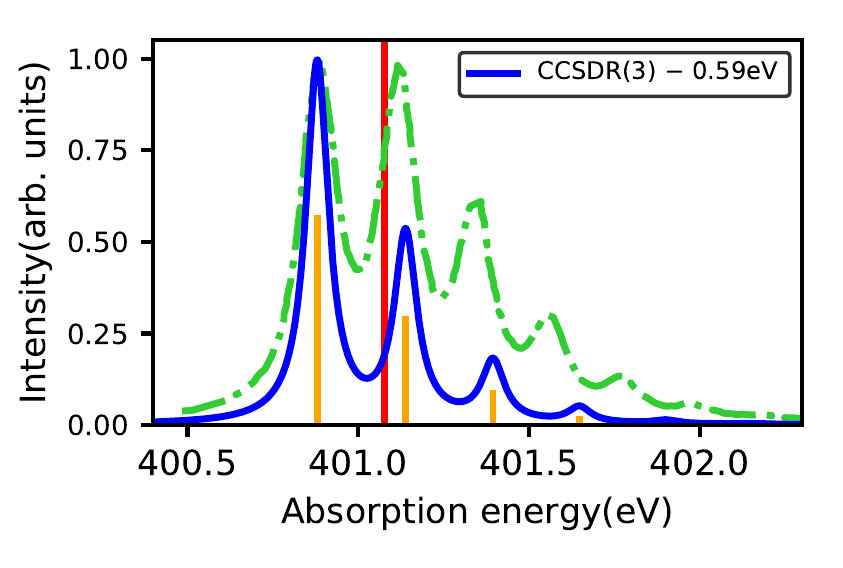}
     \end{subfigure}{}
     \begin{subfigure}[b]{0.4\textwidth}
        \centering
         \includegraphics[scale=0.8]{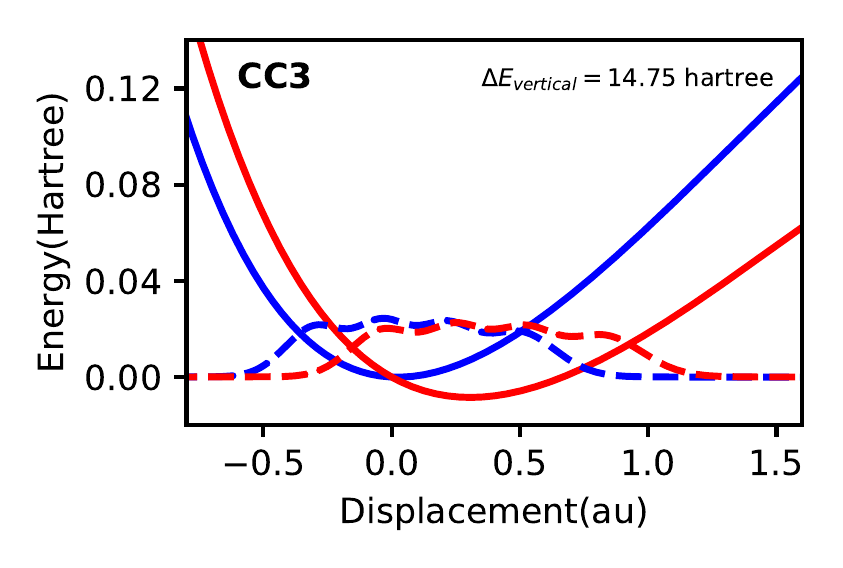}
     \end{subfigure}
     \begin{subfigure}[b]{0.4\textwidth}
      \centering
      \includegraphics[scale=0.8]{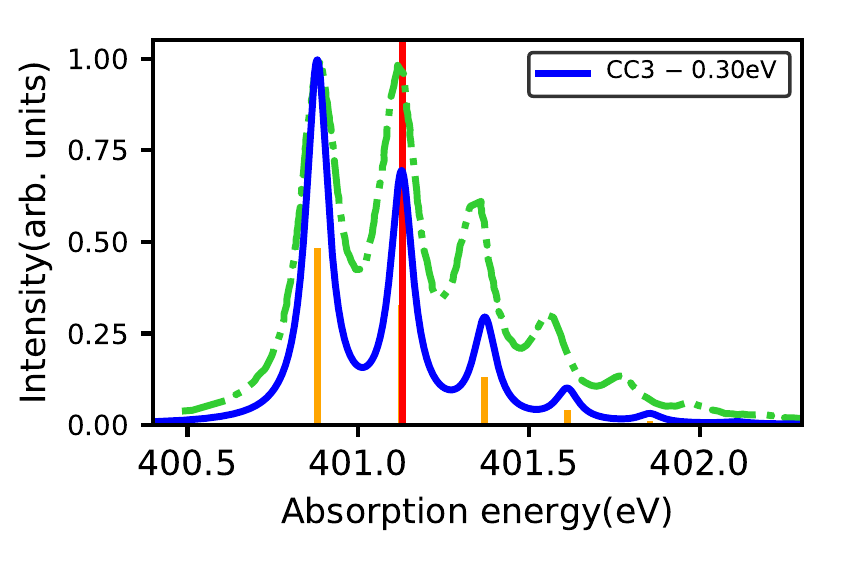}
    \end{subfigure}{}
      \caption{N$_2$: PES and vibrationally resolved XAS of the 1s$_{\rm N}$ $\rightarrow \pi^*$ transition. See the Computational Details section and the caption of Fig.~\ref{fig:C_CO} for an explanation of the notation.}
      \label{fig:N2}
\end{figure}
A combined theoretical and experimental study of core-excited states of molecular nitrogen was recently carried out by \citeauthor{Myhre_N2_2018}~\cite{Myhre_N2_2018} 
{For the 1s$\to \pi^*$ electronic transition,
the newly recorded XAS spectrum features a progression of seven peaks.}
The {computational analysis of \citeauthor{Myhre_N2_2018}}
showed that large basis sets 
were necessary for a proper {rendering} of the core {spectrum}. 
It also revealed that CC3 was capable of predicting the spectra with considerable accuracy and to assign {many of the vibrationally resolved} peaks to {specific vibronic transitions}. 
%
%
However, CCSDTQ was found to be essential to accurately reproduce the vibrational spectrum,
{and in particular the whole progression of seven peaks in the 1s$\to \pi^*$ band.}
\citeauthor{Myhre_N2_2018} stated that nitrogen is a special case with strongly interacting core holes which complicates the description of core relaxation. 

Our simulated PES and XAS of N$_2$ is shown in Fig. \ref{fig:N2}.
The nature of the PES and 
spectra are quite similar for CCSD, CCSDR(3) and CC3, showing a progression of five peaks of rapidly decreasing intensity. However, the peak separations are in reasonable agreement with the experiment. CC2 gives a considerably different result with the PES being much shallower and flared in the positive displacement region and the spectra featuring 2 extra peaks and  smaller energy gap between the peaks. 
For the CC2 spectra, we have aligned the 0-0 band with the first experimental peak instead of with most intense one. 

\subsection{Carbonyl sulfide (OCS)}
\begin{figure}[htb!]
    \centering
      \begin{subfigure}[b]{0.4\textwidth}
    \includegraphics[scale=0.8]{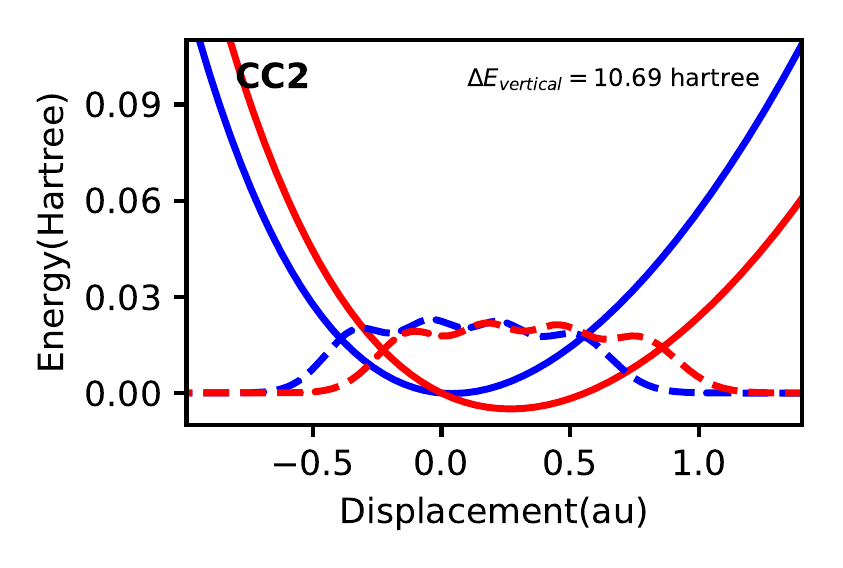}
    \end{subfigure}
    \begin{subfigure}[b]{0.4\textwidth}
    \includegraphics[scale=0.8]{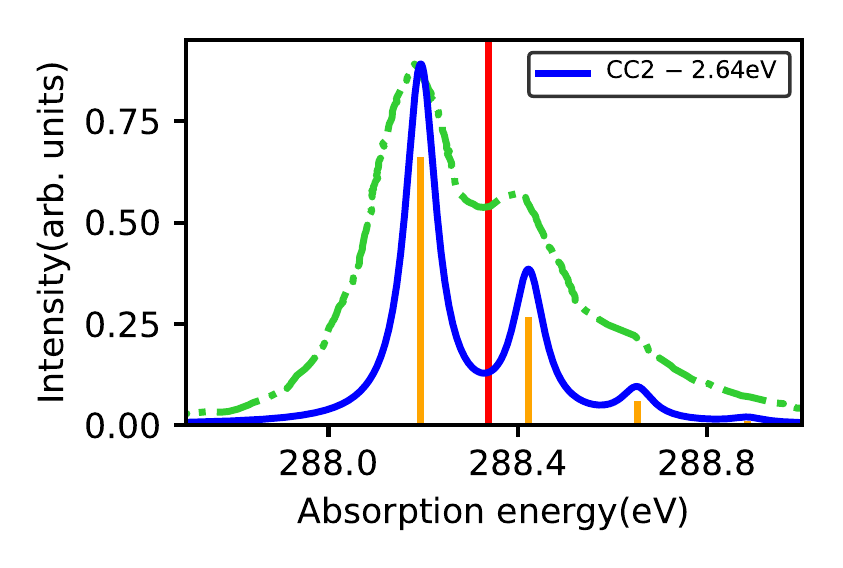}
    \end{subfigure}
      \begin{subfigure}[b]{0.4\textwidth}
    \includegraphics[scale=0.8]{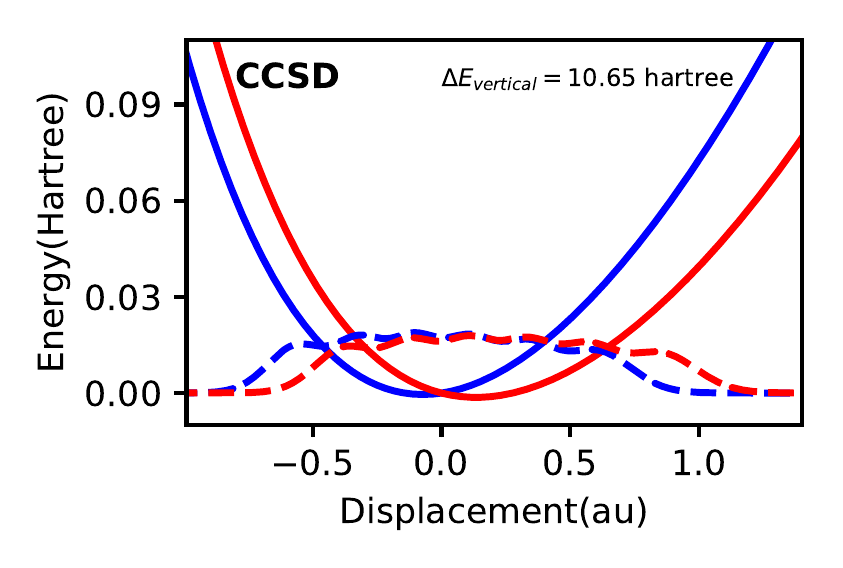}
    \end{subfigure}
    \begin{subfigure}[b]{0.4\textwidth}
    \includegraphics[scale=0.8]{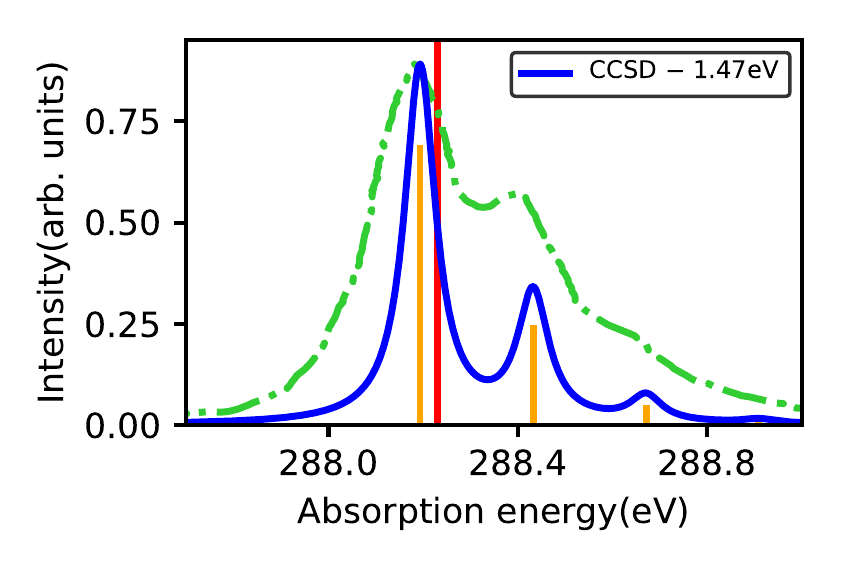}
    \end{subfigure}
    \begin{subfigure}[b]{0.4\textwidth}
    \includegraphics[scale=0.8]{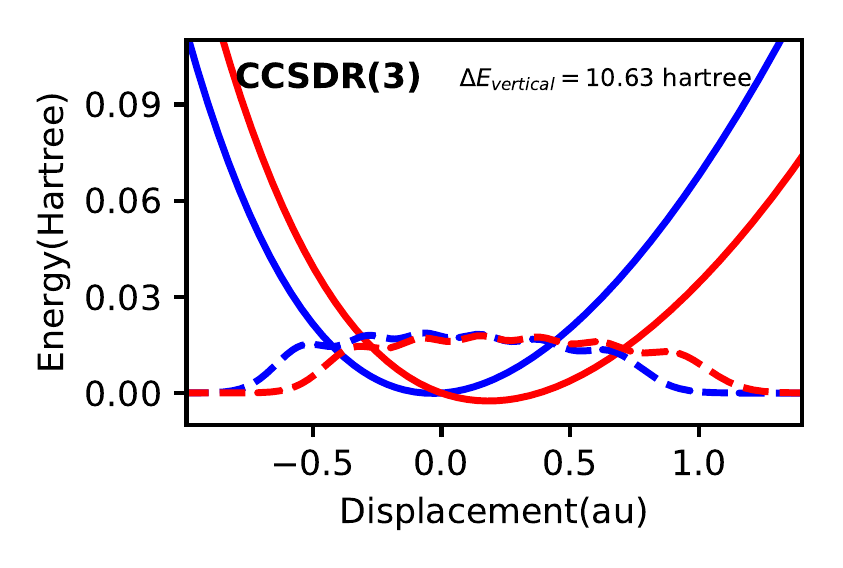}
    \end{subfigure}
    \begin{subfigure}[b]{0.4\textwidth}
    \includegraphics[scale=0.8]{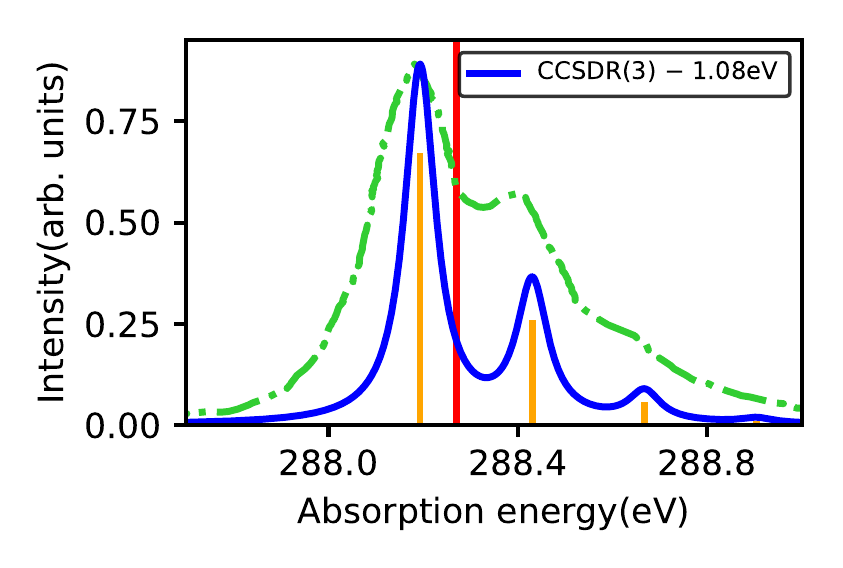}
    \end{subfigure}
    \caption{OCS: 1s$_{\rm{C}}\to\pi^*$ PES and vibrationally resolved XAS only including the CO stretching mode.}
    \label{fig:OCS}
\end{figure}
OCS is a typical molecule showing Renner-Teller effect,\cite{doi:OCS-renner} a hallmark of the breakdown of the Born-Oppenheimer approximation.\cite{renner-c2h2cation-jcp2015,renner-jcp2009} It is a signature of bending in the excited state geometry of an otherwise linear  polyatomic molecule in the ground state.\cite{Tronc_1979} {This is also confirmed in our calculations.} In the ground state, OCS has four vibrational modes, namely, a degenerate pair of O-C-S bending ($\nu_1$ and $\nu_2$), a C-S stretching ($\nu_3$) and a C-O stretching ($\nu_4$).

The experimental XAS at the C K-edge of OCS shows a vibrational progression for the 1s$_{\rm{C}} \to \pi^*$ transition, with 2 peaks and a shoulder with about 0.21 eV spacing between them.\citep{doi:OCS-renner,Tronc_1979} 
{Following the proposition made in Ref.~\citenum{doi:OCS-renner}, that the C-O stretching mode is solely responsible for the vibrational fine structure, we have carried out a first set of calculations at the CC2, CCSD and CCSDR(3) level, see Fig.~\ref{fig:OCS},
only using the CO stretching mode}. The mean densities of the ground and excited states at CC2 level are found to be more constricted in space than those for CCSD and CCSDR(3), the latter two being very similar. Both the positions and relative intensities of the peaks are in good agreement with experiment (after the rigid shift). CCSD, however, has the vertical excitation energy closest to the 0-0 FC band. A fourth peak of very low intensity is obtained for all three methods, which may have been overshadowed in the experiment.
\begin{figure}[htb!]
    \includegraphics[width=165mm]{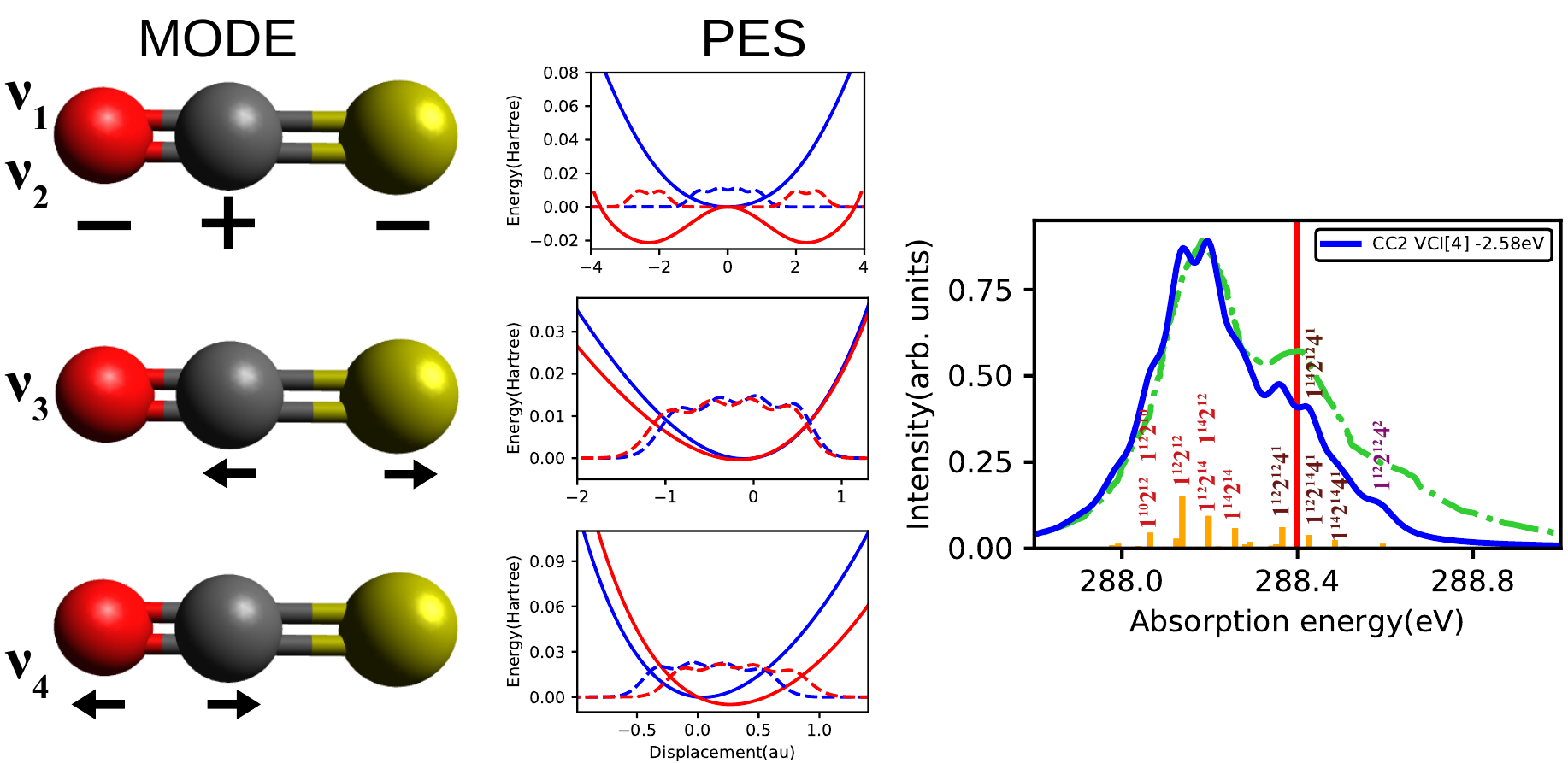}
    \caption{OCS: Ground state vibrational modes (left), where red, black and gold colors designate oxygen, carbon and sulfur atoms, respectively. 1s$_{\rm{C}}\to\pi^*$ PES (middle) constructed using the 1M ADGA scheme at the CC2/cc-pCVTZ level ($\Delta E_{vertical} =$ 10.69 hartree).  Vibrationally resolved XAS (right) generated by VCI[4] analysis. Sum of FC factor = 0.96. 
    Numeric labels indicate the vibrational mode responsible for the stick, represented as X$^y$, where X is the normal mode index and $y$ the number of quanta. 
    }
    \label{OCS-cc2-all}
\end{figure}
{In order to achieve a better match with the experimental width of each vibrational fine structure, we also performed a VCI[4] calculation of the FC factors, utilizing all four vibrational modes generated at the CC2 level using the 1M ADGA protocol. The results are shown in Fig.~\ref{OCS-cc2-all}. 
We observe that the contributions from the higher vibrational states ($v>10$) of the degenerate bending mode give richer spectral information.
The XAS can be sub-divided into regions with respect to the C-O stretching mode ($\nu_4$) progression. We show this as color blocks in Fig. \ref{OCS-cc2-all}, where red, brown and magenta labels are for $4^0$, $4^1$ and $4^2$, respectively. Multiple bending mode progressions are seen to be present in each region, explaining the broadening.
Also, it is noteworthy that in order to explicitly describe Renner-Teller effect, one should go beyond the FC approximation and include higher order terms. Yet, as a first approximation, we see here 
that FC term captures the spectral profile reasonably well.  }
The oxygen and sulphur K-edges were not considered in our study due to lack of vibrationally resolved experimental data. 
\subsection{Formaldehyde (CH$_2$O)}
\begin{table}[htbp]
    \centering
    \begin{tabular}{c|c|c|c}
     \hline
     Vibration & Symmetry & Description & Ground state frequency (cm$^{-1}$)\\
     \hline
     $\nu_1$ & B$_1$ & CH$_2$ out of plane wagging & 1210 \\
     $\nu_2$ & B$_2$ & CH$_2$ rocking & 1283\\
     $\nu_3$ & A$_1$ & CH$_2$ scissoring $+$ CO stretching & 1555\\
     $\nu_4$ & A$_1$ & CO stretching & 1778\\
     $\nu_5$ & A$_1$ & CH$_2$ symmetric stretching & 2973\\
     $\nu_6$ & B$_2$ & CH$_2$ asymmetric stretching & 3046\\
     \hline
    \end{tabular}
    \caption{CH$_2$O: All vibrational modes with frequencies in cm$^{-1}$ computed at MP2/cc-pCVTZ level.}
    \label{tab:HCHO-modes}
\end{table}
\begin{figure}[htb!]
    \centering
    \includegraphics[height=180mm]{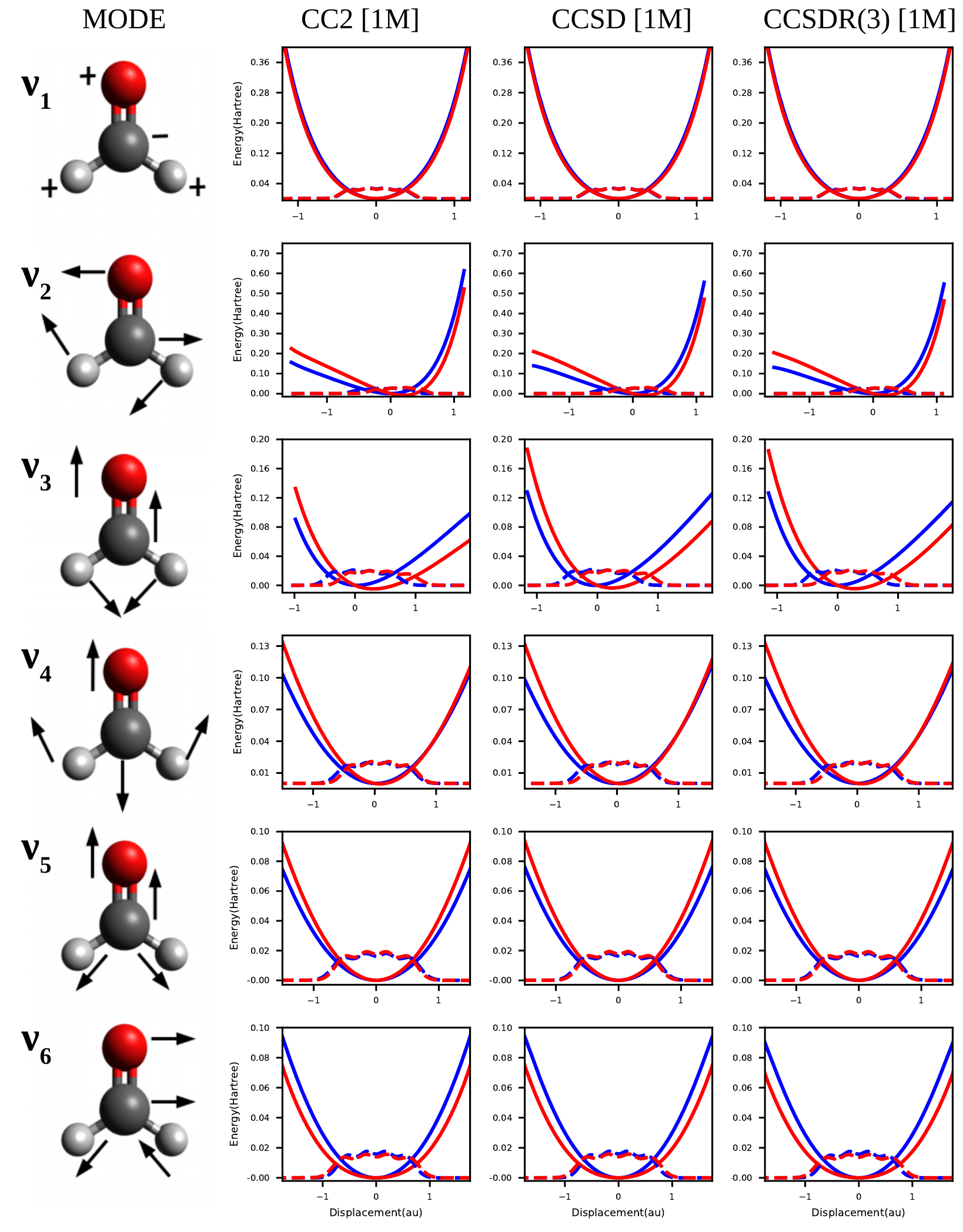}
    \caption{H$_2$CO: Vibrational modes and PES corresponding to 1s$_{\textrm{C}} \to \pi^*$ transitions generated using a 1M procedure. Red, black and grey represent oxygen, carbon and hydrogen atoms, respectively. $\Delta E_{vertical}$ is 10.60 , 10.55 and 10.53 hartree using CC2, CCSD and CCSDR(3), respectively.}
    \label{fig:C-HCHO-PES}
\end{figure}
\begin{figure}[htb!]
    \centering
    \includegraphics[scale=0.8]{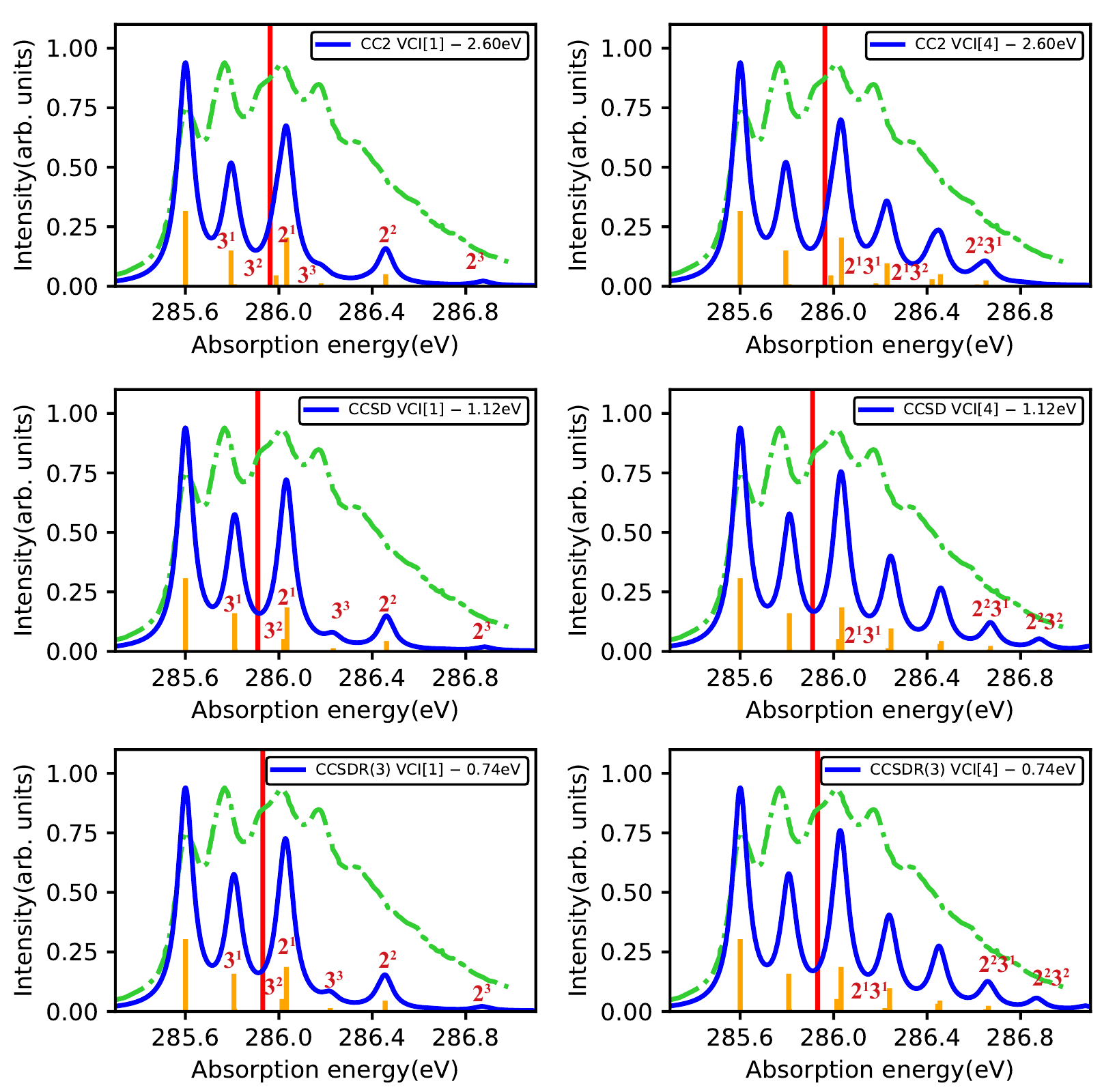}
    \caption{H$_2$CO: 1s$_{\rm{C}} \to \pi^*$ XAS {for VCI[1] and VCI[4]. The sum of FC factors at VCI[1] was found to be 0.80, 0.79 and 0.78 for CC2, CCSD and CCSDR(3)}, respectively. 1M PES are used for FC analysis. Red labels indicate the vibrational mode responsible for the peak, represented as X$^y$, where X is the normal mode index and $y$ the number of quanta. The same notation is used henceforth. }
    \label{fig:C-HCHO-XAS}
\end{figure}
\begin{figure}[htb!]
    \centering
    \includegraphics[scale=0.8]{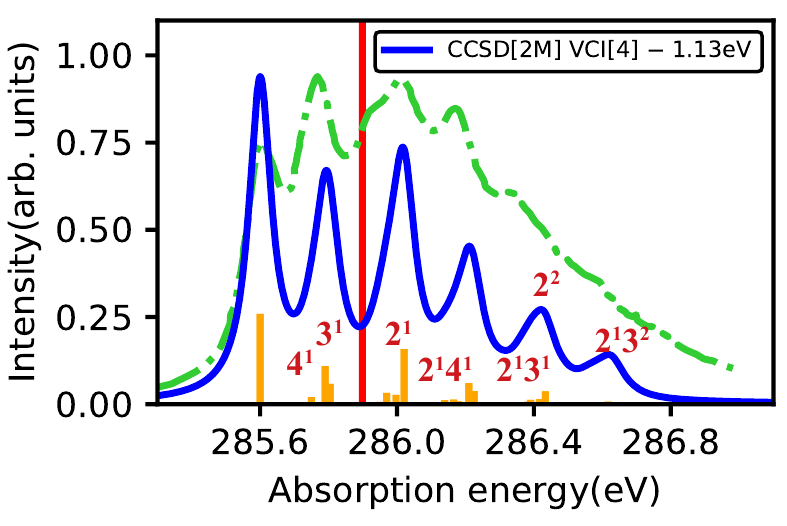}
    \caption{H$_2$CO: 1s$_{\rm{C}} \rightarrow \pi^*$ XAS using 2M PES at CCSD level of theory followed by VCI[4] analysis. The sum of FC factors is 0.97.}
    \label{fig:C_HCHO_2cut_XAS}
\end{figure}
The vibrationally resolved carbon K-edge XAS cross-section of formaldehyde has been reported by \citeauthor{remmers1992high}~\cite{remmers1992high} Between 285 and 287 eV, the 1s$_{\rm{C}} \to \pi^*$ transition was observed. The first peak corresponding to the 0-0 transition is at 285.59 eV. From an 
analysis of the fine structure, \citeauthor{remmers1992high} concluded that the normal modes corresponding to C-H and C-O stretching contributed more strongly to the spectrum than the HCH bending mode. 
Theoretical simulations have been performed before employing various electronic structure theory methods like ADC(2), MRCI and CC.\cite{formaldehyde_adc, federica2019}  
In a recent work, Frati \textit{et al.}~\cite{federica2019} used harmonic approximation adopting the AH model and computed  the ground and excited state minima with CCSD and EOM-CCSD, respectively. Their study reveals that strong Duschinsky mixing occurs between normal modes, which controls the spectral lineshape. Also, they suggest that inclusion of the effect of triple excitations is necessary to account for the elongation of core-excited state bond length. 

In our calculations, we consider all six vibrational modes (see Table \ref{tab:HCHO-modes} and Fig.~\ref{fig:C-HCHO-PES}) for all the CC methods under study. No significant differences are noticed.
Though the {separation between} 
the peaks matches reasonably with experiment (see Fig.~\ref{fig:C-HCHO-XAS}), the intensity of the 0-0 band is overestimated, leading to an inaccurate description of the spectral band shape. The sum of the Franck-Condon factor is far from unity in the VCI[1] approach (left panels of Fig. \ref{fig:C-HCHO-XAS}), indicating the necessity to include more states.
We do so using VCI[4] on top of 
1M PES
and see (right panels of Fig.~\ref{fig:C-HCHO-XAS}) the occurrence of peaks due to 2$^1$3$^1$, 2$^1$3$^2$ at CC2, CCSD and CCSDR(3) level, plus one due to 2$^2$3$^1$ upon using CCSD and CCSDR(3). For more details on the position of peaks, see Supporting Information. 

We further carried out 2M PES calculations, involving all vibrational modes, in order to consider all possible pairwise coupling terms in the generation of the PES's. 
{The 1M part of the 2M PES's are identical}, but {far more single point energy calculations are required}. 
%
The 0-0 band in the vibrationally-resolved XAS still remains overestimated in comparison to experiment, as shown in Fig.~\ref{fig:C_HCHO_2cut_XAS}. 
Additional peaks are observed which are characterized as overtones of $\nu_4$.
These do not modify the overall spectral profile when compared to the VCI[4] analysis done on the 1M PES. Also, we obtain one peak less in this calculation (observed at around {286.9} eV, in shifted scale in comparison to others), which is also reasonable as the sum of FC factor is slightly lower than unity. However, as no sharp experimental feature is observed in that region, we do not make further computational efforts to obtain it.
\begin{figure}[htb!]
    \centering
    \includegraphics[height=180mm]{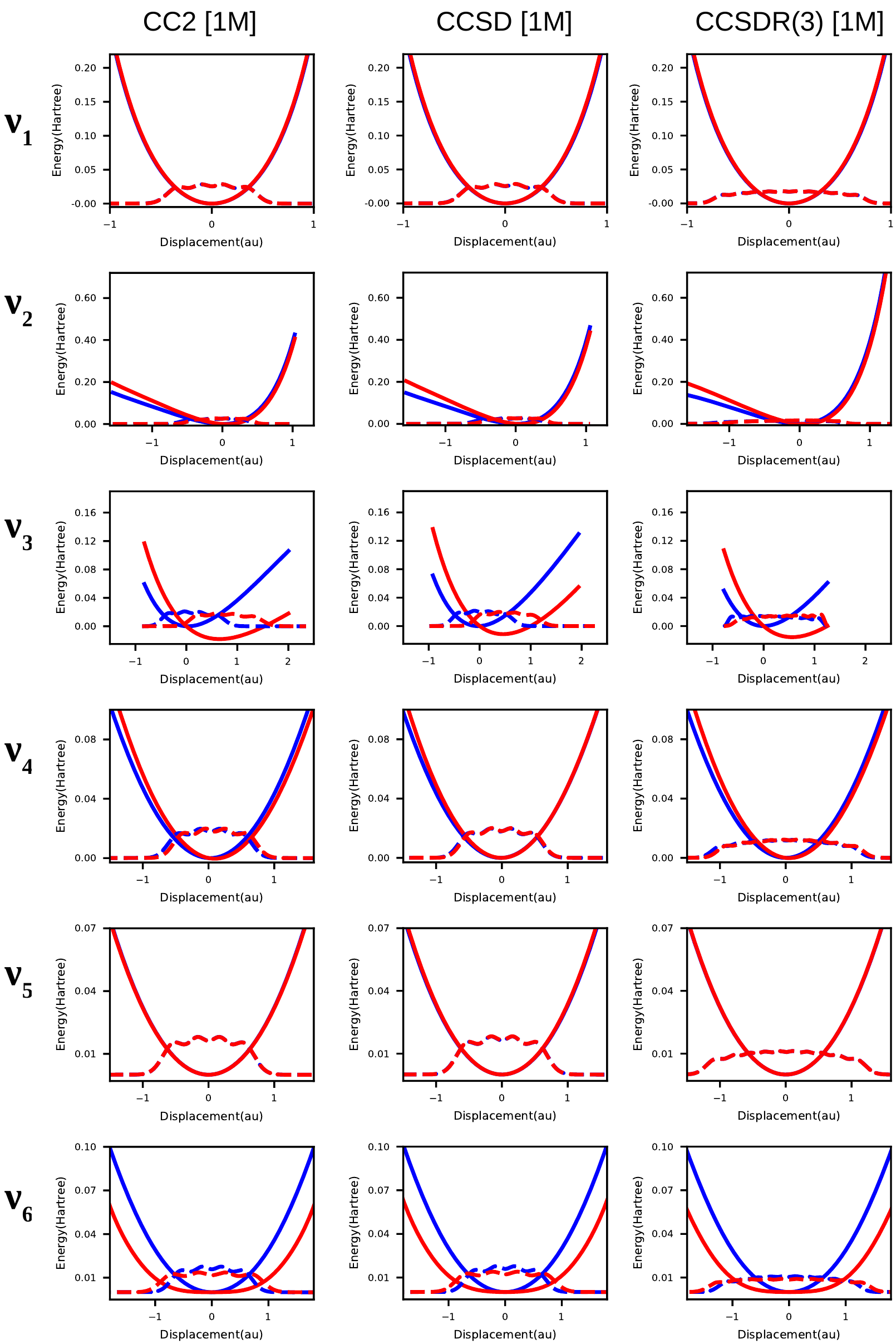}
    \caption{H$_2$CO: PESs along the 6 vibrational modes of the 1s$_{\rm O} \rightarrow \pi^*$ transition, generated using 1M approach. $\Delta E_{vertical}$ is 19.58, 19.57 and 19.53 hartree using CC2, CCSD and CCSDR(3), respectively.
    }
    \label{fig:O-HCHO-PES}
\end{figure}
\begin{figure}[htb!]
    \centering
    \includegraphics[scale=0.8]{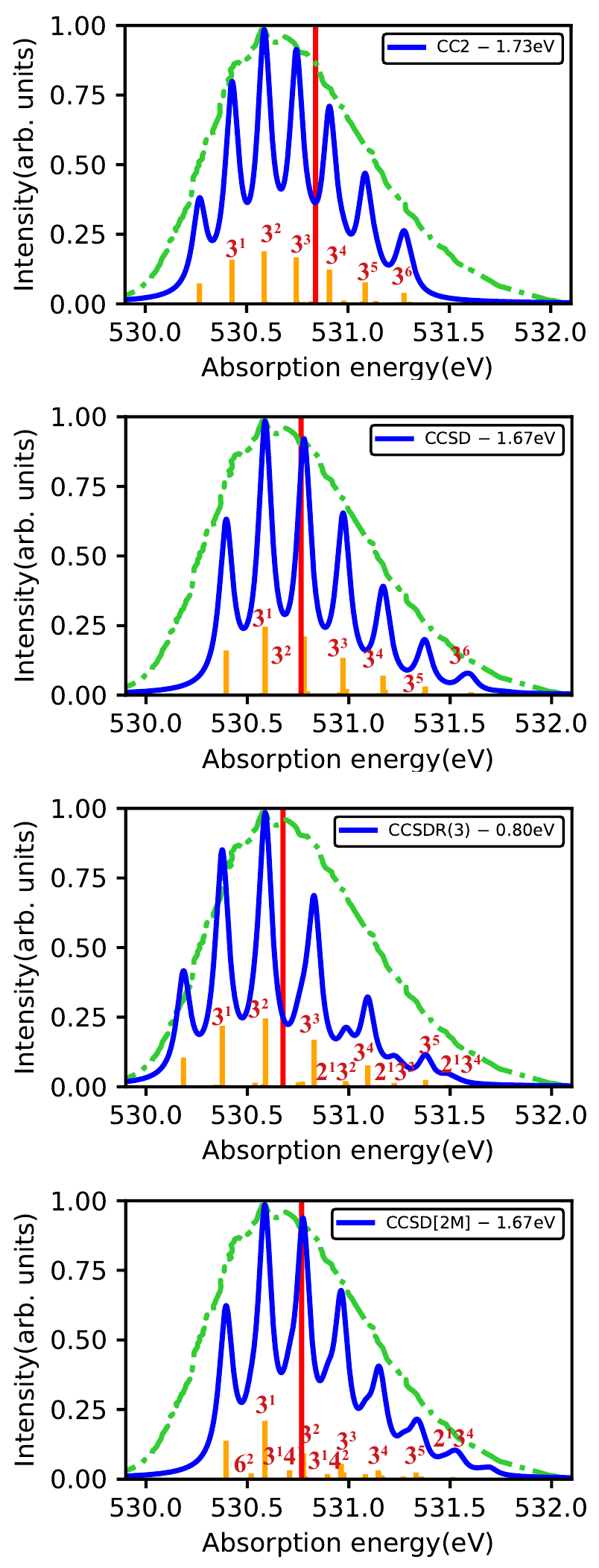}
    \caption{H$_2$CO : 1s$_{\rm{O}} \rightarrow \pi^*$ vibrationally resolved XAS using VCI[4]. }
    \label{fig:O-H2CO-XAS}
\end{figure}

Finally, the same procedure (both 1M and 2M) was applied at the oxygen K-edge, and the PES's and vibrational resolved spectra are shown in Figs.~\ref{fig:O-HCHO-PES} and ~\ref{fig:O-H2CO-XAS}, respectively.
For $\nu_1$ and $\nu_5$ the ground and excited state PESs overlap with one another completely.
We found that $\nu_3$ has a predominant role in the vibrational fine structure of the 1s$_{\rm O} \rightarrow \pi^*$ band. The experimental result was merely an envelope over the vibrational peaks and was attributed to C-O stretching motion.~\cite{Prince_small_2003} Coupling between normal modes seems to play a very minimal role as the peaks mostly have pure character. Only at the CCSDR(3) level, we observe some amount of contribution from mixed modes. 

Similar to the C K-edge results, performing 2M PES generation does not improve the spectral profile significantly, and yet the procedure is computationally more expensive. A low intensity $6^2$ spectral band appears at around 530.5 eV (in shifted scale) which is not obtained using 1M PES.

\subsection{Acetaldehyde (CH$_3$CHO)}
\begin{figure}[htb!]
        \centering
    \includegraphics[scale=0.8]{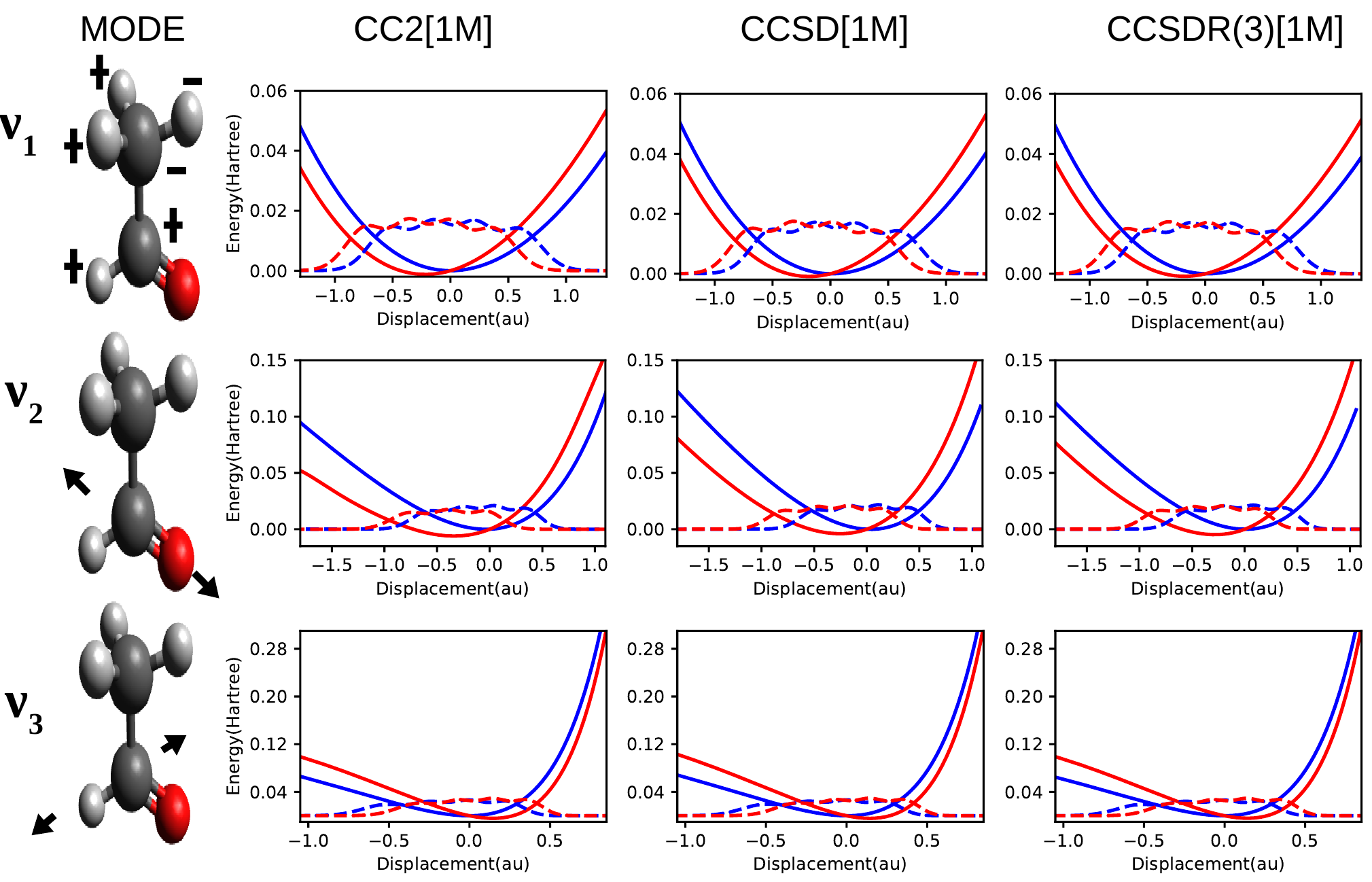}
    \caption{CH$_3$CHO: Vibrational modes and corresponding PESs for the GS and the 1s$_{\rm C} \to \pi^*$ transition, generated using 1M approach. Red, black and grey spheres represent oxygen, carbon and hydrogen atoms, respectively.
    $\Delta E_{vertical}$ is  10.62, 10.56 and 10.55 hartree using CC2, CCSD and CCSDR(3), respectively.}
    \label{fig:ch3cho-PES}
\end{figure}
\begin{figure}[htb!]
    \centering
    \includegraphics[scale=0.8]{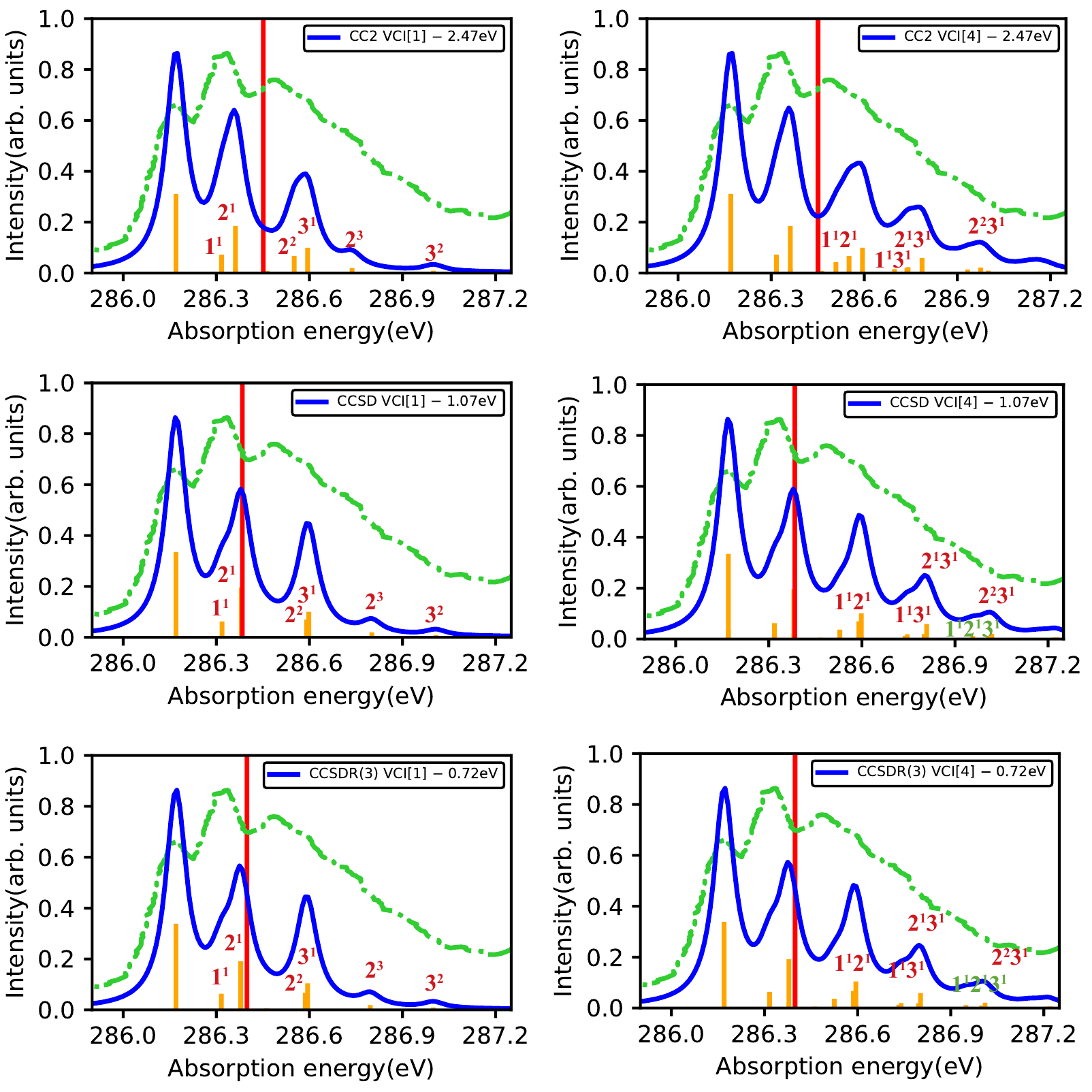}
    \caption{CH$_3$CHO: Vibrationally resolved 1s$_{\rm C} \to \pi^*$ transition. The sticks labels already marked in the VCI[1] spectra (left panel) are not repeated on the VCI[4] spectra (right panel). 1M PES is used for FC analysis. See text for further details on the notation.}
    \label{fig:ch3cho-XAS}
\end{figure}
\begin{figure}[htb!]
    \centering
    \includegraphics[scale=0.8]{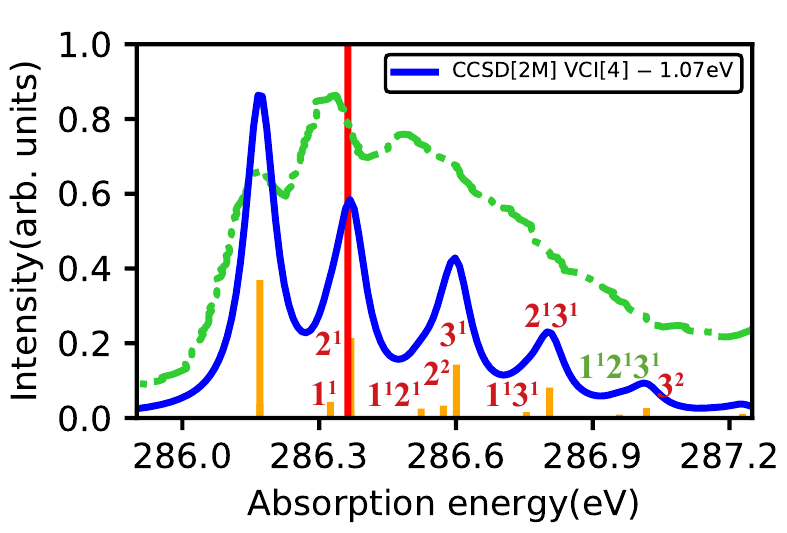}
    \caption{CH$_3$CHO: 1s$_{\rm{C}} \rightarrow \pi^*$ XAS using 2M PES at CCSD level of theory followed by VCI[4] analysis. The sum of FC factors is 0.99.}
    \label{fig:ch3cho-2cut-XAS}
\end{figure}
The carbon K-edge XAS of CH$_3$CHO consists of excitations from both the methyl and carbonyl 1s$_{\rm C}$ orbitals. The first peak centered at 286.17~eV,  due to 1s$_{\textrm{C(CO)}} \rightarrow \pi^*$, shows vibrational progression which was ascribed as originating mainly from the C-O stretching mode.\cite{Prince_small_2003}
We carried out an analysis with a smaller cc-pVDZ basis set at CCSD level using all vibrational modes to prescreen the vibrational modes having considerable contribution to the splitting. 
This revealed that major contributions 
come from the C-O stretching motion (1788 cm$^{-1}$ - $\nu_{2}$). The bending mode (1144 cm$^{-1}$ - $\nu_{1}$) and C-H stretch of the HCO moiety (2957 cm$^{-1}$ - $\nu_{3}$) also contribute to the spectral splitting. In order to strike a balance between computational cost and accuracy, 
we carried out further calculations using the cc-pCVTZ basis set considering only the 3 above mentioned modes.

The PES and the appearance of the spectrum do not change much upon moving from CC2 to CCSDR(3). Only the relative shift with respect to the 1st peak decreases as we go up the CC hierarchy. 
As seen previously {for formaldehyde}, the 0-0 band is substantially overestimated and appears to be the most intense peak in our calculation (Fig. \ref{fig:ch3cho-XAS}). In VCI[1] FC calculations, the sum of the FC factors deviates from unity, being 0.77, 0.79 and 0.80 for CC2, CCSD and CCSDR(3), respectively. This is improved upon by using the VCI[4] method. Here, we also observe a stick at 286.95 eV (in shifted scale) for CCSD and CCSDR(3) levels, (green label) involving the coupling between all the 3 modes. 

In order to consider 2M coupling while generating the PES, we consider the same 3 modes. This does not bring about noticeable change in the PES and the spectral profile. {The overestimation of the intensity of the 0-0 band and other inaccuracies suggest significant limitations {in accuracy
of the methodology used
to generate 
the excited state PES. This} 
can be due both to lack of inclusion of full triples or higher-order excitations in the electronic structure and {to limitations}
in the PES model used.
{Only considering (as done here)} a three-mode model is an approximation that may certainly have some effects. 
Note that the neglect of most modes in itself also restricts the couplings between all modes and, thereby, most pair couplings are missing. 
{We here used} one fixed set of coordinates for both the ground and excited state. The pair mode coupling terms counteract the consequences of this  {choice} by having {some of the same effects} as would be attained by considering different normal coordinates for the two states {since the wave function will describe these couplings}.}
However, a more complete PES computation at high electronic structure level was deemed disproportionately expensive at this stage.  
\section{Ethylene (C$_2$H$_4$)}
\begin{figure}[htb!]
    \centering
    \includegraphics[width=150mm]{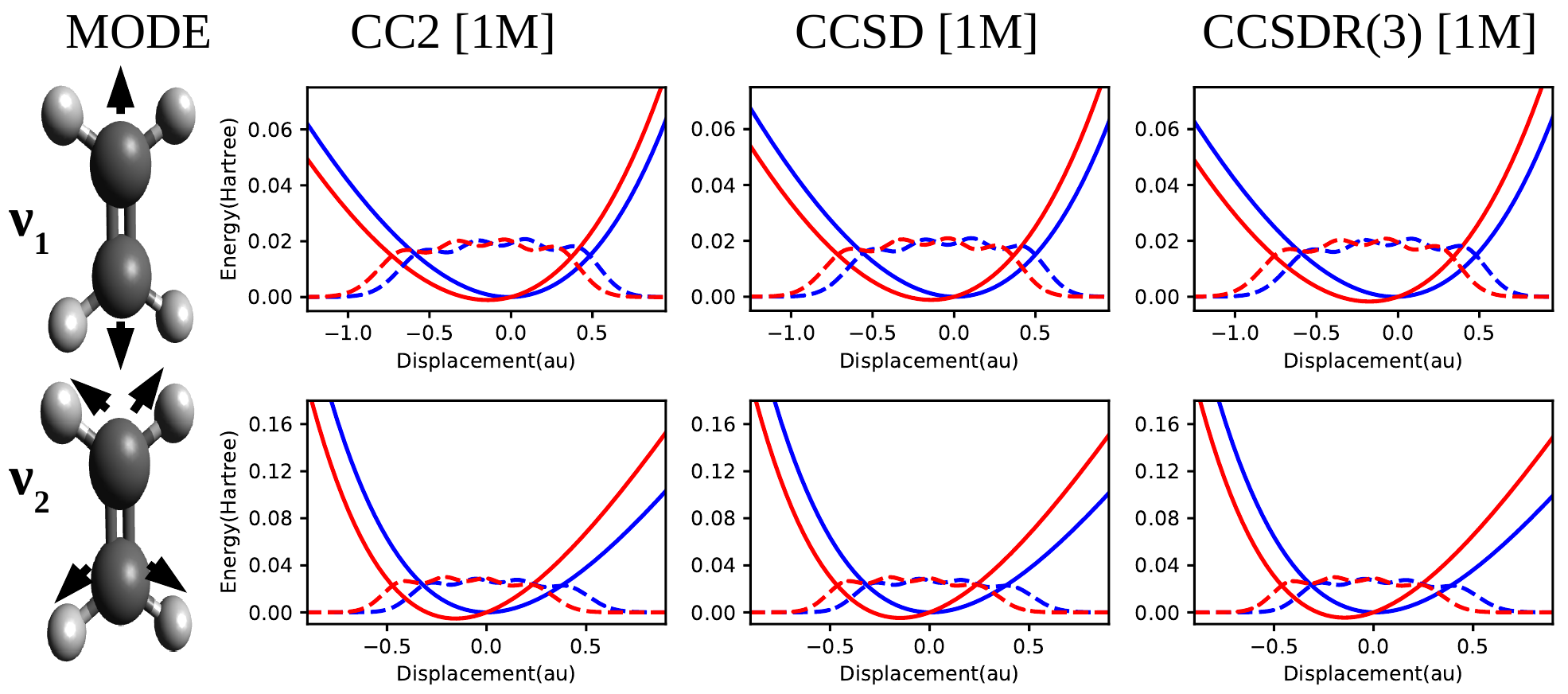}
    \caption{C$_2$H$_4$: PESs of the GS and 1s$\rightarrow\pi^*$ excited state along the in-plane C-C stretching ($\nu_1$) and C-H symmetric stretching ($\nu_2$) modes, generated using 1M approach. Black and grey spheres represent carbon and hydrogen, respectively. $\Delta E_{vertical}$ is 10.57, 10.51 and 10.49 hartree using CC2, CCSD and CCSDR(3) methods, respectively.}
    \label{fig:c2h4-mode-pes}
\end{figure}
The main peak in the carbon K-edge XAS of ethylene was found experimentally at 284.67 eV.~\cite{Tronc_1979} 
It is vibrationally resolved with a spacing of about 0.36 eV and was assigned to be of 1s$\rightarrow\pi^*$ character.
 {Our pre-screening} {protocol} at CCSD/cc-pVDZ level revealed that only 2 modes contribute towards the fine structure of the spectrum, namely, C-C stretching ($\nu_1$: 1689~cm$^{-1}$) and symmetric C-H stretching ($\nu_2$: 3201 cm$^{-1}$), i.e. two of the three total-symmetric modes.
{We then generate the 1M PES using only these 2 modes, at the various coupled cluster levels of theory with cc-pCVTZ basis set (shown in Fig. \ref{fig:c2h4-mode-pes}). The PES does not change much upon moving towards more accurate electronic structure method.   }
\begin{figure}[htb!]
    \centering
    \includegraphics[scale=0.8]{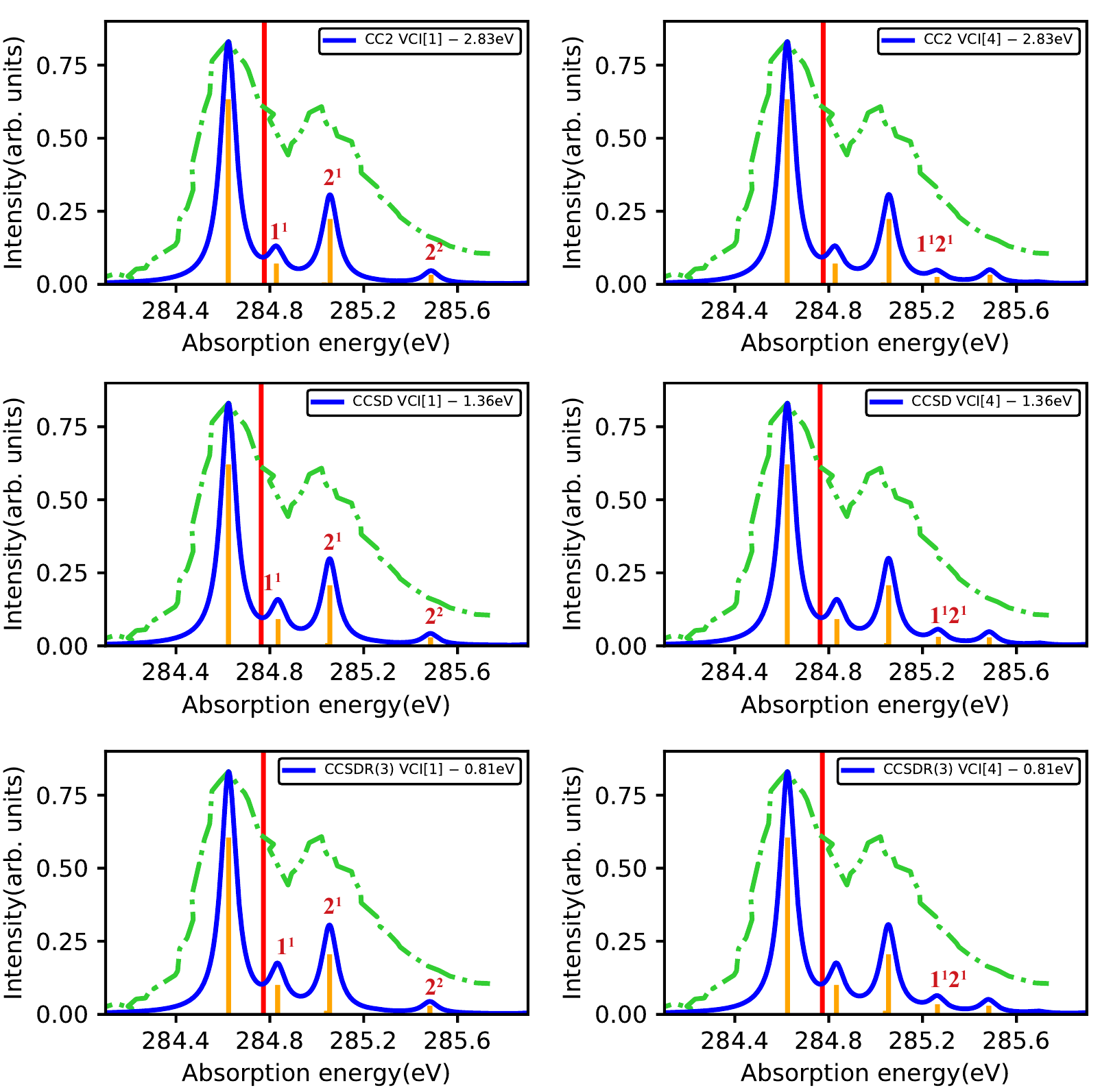}
    \caption{C$_2$H$_4$ : 1s$_{\rm C} \to \pi^*$ vibrationally resolved XAS. 1M approach is used for FC analysis. See text for details on the notation.}
    \label{fig:c2h4-xas}
\end{figure}

{With reference to the results shown in Figure
\ref{fig:c2h4-xas},}
most of the intensity is captured by VCI[1], as indicated by a sum of FC factors of about 0.9. Only an additional peak coming from the combination of the 2 modes appear at about 285.3 eV (in shifted scale) on moving to VCI[4]. 
{Here we do not carry out 2M PES calculations as 
the spectral profile is somewhat well described by 1M PES already.}



\section{Conclusion}
We have carried out an analysis of the vibrational fine structure of selected core excitations in the molecules CO, N$_2$, OCS, CH$_2$O, C$_2$H$_2$
and CH$_3$CHO, based on the ADGA procedure for the PES generation and VCI[$n$] for the determination of the FC factors. A hierarchy of CC methods was used for the PES calculations.

{As observed in previous studies, the energy shift required to align the computed spectra with the experimental one decreases as we go higher up in the CC hierarchy. 
The relative energy separations between the 
peaks at CC2 level are in good agreement with experiment, but the intensities are often inaccurate}. 
Also, CCSDR(3) improves the PES slightly over CCSD. 
CC2 and CC3 completely fail to generate the excited state PES's for the 1s$\to \pi^*$ and the 1s$\to$3s Rydberg states
at the oxygen K-edge of CO and predict them to be dissociative in nature. For polyatomic systems, 
CC2 is a reasonable bargain between accuracy and computational cost. 

A surprisingly accurate description of the vibrationally resolved XAS of the OCS molecule is captured by a pragmatic 1M approach. We observe that even though the C-O stretching mode is dominantly involved in the vibrational fine structure, the degenerate bending modes give rise to the broadening of the individual peaks. 

{In general, for non-linear polyatomic molecules, mode coupling effects in Franck Condon analysis 
is expected to play a crucial role in describing the detailed vibrational fine structure.
Nonetheless, a somewhat reasonable impression of the band-shapes and the modes and states involved 
could here be obtained using only 1M PES or 
few mode models. }

A drawback of our methodology is the inability to correctly describe the vibrational progression of the main electronic transition at the C K-edge of formaldehyde and acetaldehyde. The intensity of 0-0 band is seen to be overestimated at both 1M and 2M PES generations {even} at highest CC level followed by VCI[4] analysis. Inclusion of pair-wise coupling terms (2M) in the ADGA method for PES generation does not significantly improve the computed spectra, when considering only a selected number of modes, thus calling for full inclusion of all modes.  
{The simulated spectral profiles} 
may be improved upon by including triples or higher excitations for single point energy calculations, and/or more complete PES expansions. The computational cost will, however, increase quite significantly as a consequence of both. Other cases where 2M PES plays a noteworthy role in describing the XAS will be part of upcoming work.


In conclusion, our methodology can be used to compute the position of the vibrationally resolved bands with good precision. 
However, the fine requirements for the accurate rendering of the spectral profile, i.e. the intensities of each peak, varies from case to case. This is {perhaps} not surprising, given the different types and sizes of 
structural changes upon excitation and thus on the ability of the chosen methods to account for such changes.

\section{Acknowledgement}
T. M. and S. C. acknowledge 
support from the Marie
Sk{\l}odowska-Curie European Training Network ``COSINE - COmputational Spectroscopy In Natural sciences and Engineering'', Grant Agreement No. 765739. 
O.C. acknowledges support from the Danish Council for Independent Research through a Sapere Aude III grant (DFF - 4002-00015)

\section*{Data availability}
All data strictly necessary to reproduce the findings of this study are {incorporated in the article and available in the SI file}. 


%

\end{document}